\input harvmac.tex
\input epsf


\def\Title#1#2{\rightline{#1}\ifx\answ\bigans\nopagenumbers\pageno0\vskip1in
\else\pageno1\vskip.8in\fi \centerline{\titlefont #2}\vskip .5in}

\input epsf
\ifx\epsfbox\UnDeFiNeD\message{(NO epsf.tex, FIGURES WILL BE
IGNORED)}
\def\figin#1{\vskip2in}
\else\message{(FIGURES WILL BE INCLUDED)}\def\figin#1{#1}\fi
\def\ifig#1#2#3{\xdef#1{Fig.\the\figno}
\goodbreak\topinsert\figin{\centerline{#3}}%
\smallskip\centerline{\vbox{\baselineskip12pt
\advance\hsize by -1truein\noindent{\bf Fig.~\the\figno:} #2}}
\bigskip\endinsert\global\advance\figno by1}

\baselineskip14pt

\def\p{\partial}

%
%
\def\[{\left [}
\def\]{\right ]}
\def\({\left (}
\def\){\right )}

\def\p{\partial}

\def\Om{\Omega}

\def\a{\alpha}

\def\CO{{\cal O}}

\def\dt{\( {\p \over \p t} \)}

\def\rd{\dot r}
\def\td{\dot t}
\def\phd{\dot {\phi}}
\def\sads{AdS Schwarzschild}
\def\rh{r_+}
\def\len{{\cal L}}
\def\lenr{{\cal L}_{\rm{reg}}}
\def\rinf{r_{\rm{max}}}
\def\ri{r_i}
\def\tc{t_c}
\def\Ec{E_c}

\def\rs{r_*}
\def\vv{\tilde{v}}
\def\uu{\tilde{u}}

\def\half{{1 \over 2}}
\def\third{{1 \over 3}}
\def\quarter{{1 \over 4}}
\def\ft{{4 \over 3}}
\def\Rrr{(2 \rh^2 + R^2)}
\def\12{{1 \over 2}}
\def\Re{{\bf \rm Re}}
\def\Im{{\bf \rm Im}}
\def\AS5{{\rm Schw-AdS_5}}



\lref\KloschQV{
T.~Klosch and T.~Strobl,
Class.\ Quant.\ Grav.\  {\bf 13}, 2395 (1996)
[arXiv:gr-qc/9511081].
}

\lref\LeviCX{
T.~S.~Levi and S.~F.~Ross,
``Holography beyond the horizon and cosmic censorship,''
arXiv:hep-th/0304150.
}

\lref\Baker{ G.A. Baker and P.R. Graves-Morris, { \it Pad{\'e}
Approximants }, second edition, Cambridge Univ. Press,  (1995).}

\lref\IchinoseRG{ I.~Ichinose and Y.~Satoh, ``Entropies of scalar
fields on three-dimensional black holes,'' Nucl.\ Phys.\ B {\bf
447}, 340 (1995) [arXiv:hep-th/9412144].
}

\lref\ShenkerXQ{ S.~H.~Shenker, ``Another Length Scale in String
Theory?,'' arXiv:hep-th/9509132.
}

\lref\KabatCU{ D.~Kabat and P.~Pouliot, ``A Comment on Zero-brane
Quantum Mechanics,'' Phys.\ Rev.\ Lett.\  {\bf 77}, 1004 (1996)
[arXiv:hep-th/9603127].
}

\lref\DouglasYP{ M.~R.~Douglas, D.~Kabat, P.~Pouliot and
S.~H.~Shenker, ``D-branes and short distances in string theory,''
Nucl.\ Phys.\ B {\bf 485}, 85 (1997) [arXiv:hep-th/9608024].
}

\lref\GaiottoRM{ D.~Gaiotto, N.~Itzhaki and L.~Rastelli, ``Closed
strings as imaginary D-branes,'' arXiv:hep-th/0304192.
}

\lref\LambertZR{ N.~Lambert, H.~Liu and J.~Maldacena, `Closed
strings from decaying D-branes,'' arXiv:hep-th/0303139.
}

\lref\MaldacenaBW{ J.~M.~Maldacena and A.~Strominger, ``AdS(3)
black holes and a stringy exclusion principle,'' JHEP {\bf 9812},
005 (1998) [arXiv:hep-th/9804085].
}

\lref\GubserNZ{ S.~S.~Gubser, I.~R.~Klebanov and A.~A.~Tseytlin,
``Coupling constant dependence in the thermodynamics of N = 4
supersymmetric Yang-Mills theory,'' Nucl.\ Phys.\ B {\bf 534}, 202
(1998) [arXiv:hep-th/9805156].
}

\lref\GrisaruZN{ M.~T.~Grisaru, R.~C.~Myers and O.~Tafjord, ``SUSY
and Goliath,'' JHEP {\bf 0008}, 040 (2000) [arXiv:hep-th/0008015].
}

\lref\McGreevyCW{ J.~McGreevy, L.~Susskind and N.~Toumbas,
``Invasion of the giant gravitons from anti-de Sitter space,''
JHEP {\bf 0006}, 008 (2000) [arXiv:hep-th/0003075].
}

\lref\NunezEQ{ A.~Nunez and A.~O.~Starinets, ``AdS/CFT
correspondence, quasinormal modes, and thermal correlators in  N =
4 SYM,'' arXiv:hep-th/0302026.
}

\lref\realtime{For a review see M. Le Bellac, {\it Thermal Field
Theory}, Cambridge University Press (1996).}
\lref\SusskindKW{ L.~Susskind, ``The anthropic landscape of string
theory,'' arXiv:hep-th/0302219.
}

\lref\BerkoozJE{ M.~Berkooz, B.~Craps, D.~Kutasov and G.~Rajesh,
``Comments on cosmological singularities in string theory,'' JHEP
{\bf 0303}, 031 (2003) [arXiv:hep-th/0212215].
}

\lref\GiveonGE{ A.~Giveon, E.~Rabinovici and A.~Sever, ``Beyond
the singularity of the 2-D charged black hole,''
arXiv:hep-th/0305140.
}
\lref\GiveonGB{ A.~Giveon, E.~Rabinovici and A.~Sever, ``Strings
in singular time-dependent backgrounds,'' arXiv:hep-th/0305137.
}

\lref\LoukoTP{ J.~Louko, D.~Marolf and S.~F.~Ross, ``On geodesic
propagators and black hole holography,'' Phys.\ Rev.\ D {\bf 62},
044041 (2000) [arXiv:hep-th/0002111].
}

\lref\etinf{For reviews see A. Linde,
                 {\it Particle Physics and Inflationary Cosmology}
                 Harwood Academic, (1990), and  A.~H.~Guth, ``Inflation and eternal
inflation,'' Phys.\ Rept.\  {\bf 333}, 555 (2000)
[arXiv:astro-ph/0002156]}

\lref\BerryMI{ M. V. Berry,
 ``Uniform asymptotic smoothing of
Stokes's discontinuities," Proc.\ Roy.\ Soc.\ Lond., {\bf A422}, 7
(1989)~.}

\lref\BerryMII{ M. V. Berry and C. J. Howls,
 ``Hyperasymptotics for integrals
  with saddles," Proc.\ Roy.\ Soc.\ Lond., {\bf A434}, 657 (1991)~.}

\lref\GrossGK{ D.~J.~Gross and H.~Ooguri, ``Aspects of large N
gauge theory dynamics as seen by string theory,'' Phys.\ Rev.\ D
{\bf 58}, 106002 (1998) [arXiv:hep-th/9805129].
}

\lref\StephensAN{ C.~R.~Stephens, G.~'t Hooft and B.~F.~Whiting,
``Black hole evaporation without information loss,'' Class.\
Quant.\ Grav.\  {\bf 11}, 621 (1994) [arXiv:gr-qc/9310006].
}

\lref\SusskindIF{ L.~Susskind, L.~Thorlacius and J.~Uglum, ``The
Stretched horizon and black hole complementarity,'' Phys.\ Rev.\ D
{\bf 48}, 3743 (1993) [arXiv:hep-th/9306069].
}

%
\lref\KrausIV{ P.~Kraus, H.~Ooguri and S.~Shenker, ``Inside the
horizon with AdS/CFT,'' arXiv:hep-th/0212277.
}

\lref\Inami{ T.~Inami and H.~Ooguri, ``One Loop Effective
Potential In Anti-De Sitter Space,'' Prog.\ Theor.\ Phys.\  {\bf
73}, 1051 (1985).
}

\lref\Burgess{ C.~P.~Burgess and C.~A.~Lutken, ``Propagators And
Effective Potentials In Anti-De Sitter Space,'' Phys.\ Lett.\ B
{\bf 153}, 137 (1985).
}

\lref\MaldacenaHW{ J.~M.~Maldacena and H.~Ooguri, ``Strings in
AdS(3) and SL(2,R) WZW model. I,'' J.\ Math.\ Phys.\  {\bf 42},
2929 (2001) [arXiv:hep-th/0001053]; Phys.\ Rev.\ D {\bf 65},
106006 (2002) [arXiv:hep-th/0111180]; J.~M.~Maldacena, H.~Ooguri
and J.~Son, ``Strings in AdS(3) and the SL(2,R) WZW model. II:
Euclidean black
J.\ Math.\ Phys.\  {\bf 42}, 2961 (2001) [arXiv:hep-th/0005183];
J.~M.~Maldacena and H.~Ooguri, ``Strings in AdS(3) and the SL(2,R)
WZW model. III: Correlation functions,'' Phys.\ Rev.\ D {\bf 65},
106006 (2002) [arXiv:hep-th/0111180].
}

\lref\Ben{M. Berkooz, B. Craps, D. Kutasov, and G. Rajesh,
arXiv:hep-th/0212215.}

\lref\Tolley{ A.~J.~Tolley and N.~Turok, ``Quantum fields in a big
crunch / big bang spacetime,'' Phys.\ Rev.\ D {\bf 66}, 106005
(2002) [arXiv:hep-th/0204091].
}

\lref\Herzog{ C.~P.~Herzog and D.~T.~Son, ``Schwinger-Keldysh
Propagators from AdS/CFT Correspondence,'' arXiv:hep-th/0212072.
}


\lref\KutasovXU{ D.~Kutasov and N.~Seiberg, ``More comments on
string theory on AdS(3),'' JHEP {\bf 9904}, 008 (1999)
[arXiv:hep-th/9903219].
}

\lref\deBoerPP{ J.~de Boer, H.~Ooguri, H.~Robins and
J.~Tannenhauser, ``String theory on AdS(3),'' JHEP {\bf 9812}, 026
(1998) [arXiv:hep-th/9812046].
}


\lref\CruzIR{ N.~Cruz, C.~Martinez and L.~Pena, ``Geodesic
Structure Of The (2+1) Black Hole,''
[arXiv:gr-qc/9401025].
}

\lref\MartinecXQ{ E.~J.~Martinec and W.~McElgin, ``Exciting AdS
orbifolds,'' arXiv:hep-th/0206175.
}

\lref\GiveonNS{ A.~Giveon, D.~Kutasov and N.~Seiberg, ``Comments
on string theory on AdS(3),'' Adv.\ Theor.\ Math.\ Phys.\  {\bf
2}, 733 (1998) [arXiv:hep-th/9806194].
}

\lref\SonQM{ J.~Son, ``String theory on AdS(3)/Z(N),''
arXiv:hep-th/0107131.
}

\lref\HananyEV{ A.~Hanany, N.~Prezas and J.~Troost, ``The
partition function of the two-dimensional black hole conformal
JHEP {\bf 0204}, 014 (2002) [arXiv:hep-th/0202129].
}

\lref\TeschnerFT{ J.~Teschner, Nucl.\ Phys.\ B {\bf 546}, 390
(1999) [arXiv:hep-th/9712256];
%
Nucl.\ Phys.\ B {\bf 571}, 555 (2000) [arXiv:hep-th/9906215].
}

\lref\DanielssonZT{ U.~H.~Danielsson, E.~Keski-Vakkuri and
M.~Kruczenski, ``Spherically collapsing matter in AdS, holography,
and shellons,'' Nucl.\ Phys.\ B {\bf 563}, 279 (1999)
[arXiv:hep-th/9905227].
}
\lref\EvansFR{ T.~S.~Evans, A.~Gomez Nicola, R.~J.~Rivers and
D.~A.~Steer, ``Transport coefficients and analytic continuation in
dual 1+1
arXiv:hep-th/0204166.
}

\lref\BanadosWN{ M.~Banados, C.~Teitelboim and J.~Zanelli, ``The
Black Hole In Three-Dimensional Space-Time,'' Phys.\ Rev.\ Lett.\
{\bf 69}, 1849 (1992) [arXiv:hep-th/9204099];
M.~Banados, M.~Henneaux, C.~Teitelboim and J.~Zanelli, ``Geometry
of the (2+1) black hole,'' Phys.\ Rev.\ D {\bf 48}, 1506 (1993)
[arXiv:gr-qc/9302012].}

\lref\IsraelUR{ W.~Israel, ``Thermo Field Dynamics Of Black
Holes,'' Phys.\ Lett.\ A {\bf 57}, 107 (1976).
}

\lref\UnruhDB{ W.~G.~Unruh, ``Notes On Black Hole Evaporation,''
Phys.\ Rev.\ D {\bf 14}, 870 (1976).
}

\lref\ElitzurRT{ S.~Elitzur, A.~Giveon, D.~Kutasov and
E.~Rabinovici, ``From big bang to big crunch and beyond,'' JHEP
{\bf 0206}, 017 (2002) [arXiv:hep-th/0204189].
}

\lref\CrapsII{ B.~Craps, D.~Kutasov and G.~Rajesh, ``String
propagation in the presence of cosmological singularities,'' JHEP
{\bf 0206}, 053 (2002) [arXiv:hep-th/0205101].
}

\lref\CornalbaNV{ L.~Cornalba, M.~S.~Costa and C.~Kounnas, ``A
resolution of the cosmological singularity with orientifolds,''
arXiv:hep-th/0204261.
}

\lref\GubserBC{ S.~S.~Gubser, I.~R.~Klebanov and A.~M.~Polyakov,
``Gauge theory correlators from non-critical string theory,''
Phys.\ Lett.\ B {\bf 428}, 105 (1998) [arXiv:hep-th/9802109].
}

\lref\KeskiVakkuriNW{ E.~Keski-Vakkuri, ``Bulk and boundary
dynamics in BTZ black holes,'' Phys.\ Rev.\ D {\bf 59}, 104001
(1999) [arXiv:hep-th/9808037].
}

\lref\MaldacenaRE{ J.~M.~Maldacena, ``The large N limit of
superconformal field theories and supergravity,'' Adv.\ Theor.\
Math.\ Phys.\  {\bf 2}, 231 (1998) [Int.\ J.\ Theor.\ Phys.\  {\bf
38}, 1113 (1999)] [arXiv:hep-th/9711200].
}

\lref\LifschytzEB{ G.~Lifschytz and M.~Ortiz, ``Scalar Field
Quantization On The (2+1)-Dimensional Black Hole
Phys.\ Rev.\ D {\bf 49}, 1929 (1994) [arXiv:gr-qc/9310008].
}

\lref\BalasubramanianSN{ V.~Balasubramanian, P.~Kraus and
A.~E.~Lawrence, ``Bulk vs. boundary dynamics in anti-de Sitter
spacetime,'' Phys.\ Rev.\ D {\bf 59}, 046003 (1999)
[arXiv:hep-th/9805171].
}

\lref\WittenQJ{ E.~Witten, ``Anti-de Sitter space and
holography,'' Adv.\ Theor.\ Math.\ Phys.\  {\bf 2}, 253 (1998)
[arXiv:hep-th/9802150].
}

\lref\BalasubramanianRE{ V.~Balasubramanian and P.~Kraus, ``A
stress tensor for anti-de Sitter gravity,'' Commun.\ Math.\ Phys.\
{\bf 208}, 413 (1999) [arXiv:hep-th/9902121].
}

\lref\HorowitzXK{ G.~T.~Horowitz and D.~Marolf, ``A new approach
to string cosmology,'' JHEP {\bf 9807}, 014 (1998)
[arXiv:hep-th/9805207].
}

\lref\BalasubramanianDE{ V.~Balasubramanian, P.~Kraus,
A.~E.~Lawrence and S.~P.~Trivedi, ``Holographic probes of anti-de
Sitter space-times,'' Phys.\ Rev.\ D {\bf 59}, 104021 (1999)
[arXiv:hep-th/9808017].
}

\lref\CarneirodaCunhaNW{ B.~G.~Carneiro da Cunha,
``Three-dimensional de Sitter gravity and the correspondence,''
Phys.\ Rev.\ D {\bf 65}, 104025 (2002) [arXiv:hep-th/0110169].
}

\lref\MaldacenaKR{ J.~M.~Maldacena, ``Eternal black holes in
Anti-de-Sitter,'' arXiv:hep-th/0106112.
}

\lref\GiveonNS{ A.~Giveon, D.~Kutasov and N.~Seiberg, ``Comments
on string theory on AdS(3),'' Adv.\ Theor.\ Math.\ Phys.\  {\bf
2}, 733 (1998) [arXiv:hep-th/9806194].
}

\lref\TeschnerFT{ J.~Teschner, ``On structure constants and fusion
rules in the SL(2,C)/SU(2) WZNW
Nucl.\ Phys.\ B {\bf 546}, 390 (1999) [arXiv:hep-th/9712256];
``Operator product expansion and factorization in the H-3+ WZNW
Nucl.\ Phys.\ B {\bf 571}, 555 (2000) [arXiv:hep-th/9906215].
}

\lref\HorowitzJC{ G.~T.~Horowitz and D.~L.~Welch, ``Exact
three-dimensional black holes in string theory,'' Phys.\ Rev.\
Lett.\  {\bf 71}, 328 (1993) [arXiv:hep-th/9302126];
%
N.~Kaloper,
Phys.\ Rev.\ D {\bf 48}, 2598 (1993) [arXiv:hep-th/9303007];
%
M.~Natsuume and Y.~Satoh, ``String theory on three dimensional
black holes,'' Int.\ J.\ Mod.\ Phys.\ A {\bf 13}, 1229 (1998)
[arXiv:hep-th/9611041];
Y.~Satoh, ``Ghost-free and modular invariant spectra of a string
in SL(2,R) and
Nucl.\ Phys.\ B {\bf 513}, 213 (1998) [arXiv:hep-th/9705208];
%
J.~M.~Maldacena and A.~Strominger, ``AdS(3) black holes and a
stringy exclusion principle,'' JHEP {\bf 9812}, 005 (1998)
[arXiv:hep-th/9804085];
%
S.~Hemming and E.~Keski-Vakkuri,
Nucl.\ Phys.\ B {\bf 626}, 363 (2002) [arXiv:hep-th/0110252];
%
 J.~Troost,
``Winding strings and AdS(3) black holes,'' arXiv:hep-th/0206118;
%
E.~J.~Martinec and W.~McElgin, ``String theory on AdS orbifolds,''
JHEP {\bf 0204}, 029 (2002) [arXiv:hep-th/0106171];
}

\lref\BalasubramanianRY{ V.~Balasubramanian, S.~F.~Hassan,
E.~Keski-Vakkuri and A.~Naqvi, ``A space-time orbifold: A toy
model for a cosmological
arXiv:hep-th/0202187.
}

\lref\CornalbaFI{ L.~Cornalba and M.~S.~Costa, ``A New
Cosmological Scenario in String Theory,'' arXiv:hep-th/0203031.
}

\lref\NekrasovKF{ N.~Nekrasov, ``Milne universe, tachyons, and
quantum group'' arXiv: hep-th/0203112.
}

\lref\SimonMA{ J.~Simon, ``The geometry of null rotation
identifications,'' JHEP {\bf 0206}, 001 (2002)
[arXiv:hep-th/0203201].
}

\lref\LiuFT{ H.~Liu, G.~Moore and N.~Seiberg, ``Strings in a
time-dependent orbifold,'' JHEP {\bf 0206}, 045 (2002)
[arXiv:hep-th/0204168];
``Strings in time-dependent orbifolds,'' arXiv:hep-th/0206182.
}

\lref\LawrenceAJ{ A.~Lawrence, ``On the instability of 3D null
singularities,'' arXiv:hep-th/0205288.
}

\lref\FabingerKR{ M.~Fabinger and J.~McGreevy, ``On smooth
time-dependent orbifolds and null singularities,''
arXiv:hep-th/0206196.
}

\lref\HorowitzMW{ G.~T.~Horowitz and J.~Polchinski, ``Instability
of spacelike and null orbifold singularities,''
arXiv:hep-th/0206228.
}

\lref\SusskindQC{ L.~Susskind and J.~Uglum, ``String Physics and
Black Holes,'' Nucl.\ Phys.\ Proc.\ Suppl.\  {\bf 45BC}, 115
(1996) [arXiv:hep-th/9511227].
}

\lref\HartleAI{ J.~B.~Hartle and S.~W.~Hawking, ``Wave Function Of
The Universe,'' Phys.\ Rev.\ D {\bf 28}, 2960 (1983).
}

\lref\HartleTP{ J.~B.~Hartle and S.~W.~Hawking, ``Path Integral
Derivation Of Black Hole Radiance,'' Phys.\ Rev.\ D {\bf 13}, 2188
(1976).
}

\lref\NiemiNF{ A.~J.~Niemi and G.~W.~Semenoff, ``Finite
Temperature Quantum Field Theory In Minkowski Space,'' Annals
Phys.\  {\bf 152}, 105 (1984).
}

\lref\HemmingKD{ S.~Hemming, E.~Keski-Vakkuri and P.~Kraus,
``Strings in the extended BTZ spacetime,'' JHEP {\bf 0210}, 006
(2002) [arXiv:hep-th/0208003].
}

\lref\DixonJW{ L.~J.~Dixon, J.~A.~Harvey, C.~Vafa and E.~Witten,
``Strings On Orbifolds,'' Nucl.\ Phys.\ B {\bf 261}, 678 (1985).
}

\lref\StromingerCZ{ A.~Strominger, ``Massless black holes and
conifolds in string theory,'' Nucl.\ Phys.\ B {\bf 451}, 96 (1995)
[arXiv:hep-th/9504090].
}

\lref\JohnsonQT{ C.~V.~Johnson, A.~W.~Peet and J.~Polchinski,
``Gauge theory and the excision of repulson singularities,''
Phys.\ Rev.\ D {\bf 61}, 086001 (2000) [arXiv:hep-th/9911161].
}

\lref\BanksVH{ T.~Banks, W.~Fischler, S.~H.~Shenker and
L.~Susskind, ``M theory as a matrix model: A conjecture,'' Phys.\
Rev.\ D {\bf 55}, 5112 (1997) [arXiv:hep-th/9610043].
}

\lref\RastelliUV{ For a review of some recent developments see
L.~Rastelli, A.~Sen and B.~Zwiebach, ``Vacuum string field
theory,'' arXiv:hep-th/0106010.
}

\lref\WittenZW{ E.~Witten, ``Anti-de Sitter space, thermal phase
transition, and confinement in gauge theories,'' Adv.\ Theor.\
Math.\ Phys.\  {\bf 2}, 505 (1998) [arXiv:hep-th/9803131].
}

\lref\BalasubramanianZV{ V.~Balasubramanian and S.~F.~Ross,
``Holographic particle detection,'' Phys.\ Rev.\ D {\bf 61},
044007 (2000) [arXiv:hep-th/9906226].
}
\lref\HubenyDG{ V.~E.~Hubeny, ``Precursors see inside black
holes,'' arXiv:hep-th/0208047.
}

\lref\SusskindIF{ L.~Susskind, L.~Thorlacius and J.~Uglum, ``The
Stretched horizon and black hole complementarity,'' Phys.\ Rev.\ D
{\bf 48}, 3743 (1993) [arXiv:hep-th/9306069].
}

\lref\gks{N. Goheer, M. Kleban, and L. Susskind,  ``The Trouble with de
Sitter Space,"
[arXiv:hep-th/0212209].}

\lref\dks{L. Dyson, M. Kleban, and L. Susskind,  ``Disturbing Implications of a Cosmological Constant,"
JHEP 0210:011,2002 [arXiv:hep-th/0208013].

\lref\MaldacenaBW{ J.~M.~Maldacena and A.~Strominger, ``AdS(3)
black holes and a stringy exclusion principle,'' JHEP {\bf 9812},
005 (1998) [arXiv:hep-th/9804085].
}

\lref\DysonNT{ L.~Dyson, J.~Lindesay and L.~Susskind, ``Is there
really a de Sitter/CFT duality,'' JHEP {\bf 0208}, 045 (2002)
[arXiv:hep-th/0202163].}
}

\lref\WittenQJ{
E.~Witten,
``Anti-de Sitter space and holography,''
Adv.\ Theor.\ Math.\ Phys.\  {\bf 2}, 253 (1998)
[arXiv:hep-th/9802150].
}

\lref\AharonyTI{
O.~Aharony, S.~S.~Gubser, J.~M.~Maldacena, H.~Ooguri and Y.~Oz,
``Large N field theories, string theory and gravity,''
Phys.\ Rept.\  {\bf 323}, 183 (2000)
[arXiv:hep-th/9905111].
}
\lref\GiddingsPT{
S.~B.~Giddings and M.~Lippert,
``Precursors, black holes, and a locality bound,''
Phys.\ Rev.\ D {\bf 65}, 024006 (2002)
[arXiv:hep-th/0103231].
}

\lref\FreivogelEX{
B.~Freivogel, S.~B.~Giddings and M.~Lippert,
``Toward a theory of precursors,''
Phys.\ Rev.\ D {\bf 66}, 106002 (2002)
[arXiv:hep-th/0207083].
}

\lref\HorowitzFM{
G.~T.~Horowitz and V.~E.~Hubeny,
``CFT description of small objects in AdS,''
JHEP {\bf 0010}, 027 (2000)
[arXiv:hep-th/0009051].
}

\lref\JacobsonMI{
T.~Jacobson,
``On the nature of black hole entropy,''
arXiv:gr-qc/9908031.
}

\lref\BanksDD{
T.~Banks, M.~R.~Douglas, G.~T.~Horowitz and E.~J.~Martinec,
``AdS dynamics from conformal field theory,''
arXiv:hep-th/9808016.
}

\lref\BalasubramanianDE{
V.~Balasubramanian, P.~Kraus, A.~E.~Lawrence and S.~P.~Trivedi,
``Holographic probes of anti-de Sitter space-times,''
Phys.\ Rev.\ D {\bf 59}, 104021 (1999)
[arXiv:hep-th/9808017].
}

\lref\KabatYQ{
D.~Kabat and G.~Lifschytz,
``Gauge theory origins of supergravity causal structure,''
JHEP {\bf 9905}, 005 (1999)
[arXiv:hep-th/9902073].
}

\lref\DanielssonFA{
U.~H.~Danielsson, E.~Keski-Vakkuri and M.~Kruczenski,
``Black hole formation in AdS and thermalization on the boundary,''
JHEP {\bf 0002}, 039 (2000)
[arXiv:hep-th/9912209].
}

\lref\GregoryAN{
J.~P.~Gregory and S.~F.~Ross,
``Looking for event horizons using UV/IR relations,''
Phys.\ Rev.\ D {\bf 63}, 104023 (2001)
[arXiv:hep-th/0012135].
}

\lref\SusskindEY{
L.~Susskind and N.~Toumbas,
``Wilson loops as precursors,''
Phys.\ Rev.\ D {\bf 61}, 044001 (2000)
[arXiv:hep-th/9909013].
}

\lref\HashimotoZP{
A.~Hashimoto, S.~Hirano and N.~Itzhaki,
``Large branes in AdS and their field theory dual,''
JHEP {\bf 0008}, 051 (2000)
[arXiv:hep-th/0008016].
}

\lref\HawkingDH{
S.~W.~Hawking and D.~N.~Page,
``Thermodynamics Of Black Holes In Anti-De Sitter Space,''
Commun.\ Math.\ Phys.\  {\bf 87}, 577 (1983).
}

\lref\SusskindVU{ L.~Susskind, ``The World as a hologram,'' J.\
Math.\ Phys.\  {\bf 36}, 6377 (1995) [arXiv:hep-th/9409089].
}

\lref\GubserBC{ S.~S.~Gubser, I.~R.~Klebanov and A.~M.~Polyakov,
``Gauge theory correlators from non-critical string theory,''
Phys.\ Lett.\ B {\bf 428}, 105 (1998) [arXiv:hep-th/9802109].
}

\Title{\vbox{\baselineskip12pt \hbox{hep-th/0306170}
\hbox{SU-ITP-03-16} \vskip-.4in}}  {The Black Hole Singularity in
AdS/CFT }

\centerline{Lukasz Fidkowski, Veronika Hubeny, Matthew Kleban, and
Stephen Shenker}
\bigskip

\centerline{\it Department of Physics, Stanford University,
Stanford, CA, 94305, USA}

\bigskip\bigskip \baselineskip14pt

\noindent

We explore physics behind the horizon in eternal AdS Schwarzschild
black holes.   In dimension $d >3$ ,
where the curvature grows large near the singularity, we find
distinct but subtle signals of this singularity in the boundary
CFT correlators. Building on previous work, we study correlation
functions of operators on the two disjoint asymptotic boundaries
of the spacetime by investigating the spacelike geodesics that
join the boundaries. These dominate the correlators for large mass
bulk fields.  We show that the Penrose diagram for $d>3$
is not square.  As a result, the real geodesic connecting the two boundary points becomes
almost null and bounces off the singularity at a finite boundary time $t_c \neq
0$. 

If this geodesic were to dominate the correlator there would be a
``light cone" singularity at $t_c$.  However, general properties of the
boundary theory rule this out. In fact, we argue that the
correlator is actually dominated by a complexified geodesic, whose properties yield
 the large mass quasinormal mode frequencies previously
found for this black hole.
We find a branch cut in the correlator at small time
(in the limit of large mass), arising from coincidence
of three geodesics.
The $t_c$ singularity, a signal of the black hole singularity, occurs on a
secondary sheet of the analytically continued correlator.  Its properties are 
computationally accessible.  

The $t_c$ singularity persists to all orders in the $1/m$
expansion, for finite $\alpha'$, and to all orders in $g_s$.
Certain leading nonperturbative effects can also be studied. The
behavior of these boundary theory quantities near $t_c$ gives, in
principle, significant information about stringy and quantum
behavior in the vicinity of the black hole singularity.

\Date{June, 2003}

\newsec{Introduction}

Some of the deepest mysteries in quantum gravity lie hidden behind
horizons, including the nature of the spacelike singularities
inside black holes.  A better understanding of these may shed more
light on cosmological singularities as well.   Despite much recent
work \refs{\BalasubramanianRY,\CornalbaFI,\NekrasovKF,\SimonMA,
\LiuFT,\ElitzurRT,\CornalbaNV,\CrapsII,\FabingerKR,
\LawrenceAJ,\HorowitzMW, \BerkoozJE, \GiveonGE, \GiveonGB }, these
issues remain mysterious.  More generally, the existence of
horizons and resolving the puzzles of information loss and quantum
decoherence require that the description of physics inside the
horizon be linked somehow to the description outside.   This
connection--the central concept of black hole complementarity
\refs{\StephensAN, \SusskindIF}--is still not well understood.

As observations continue to bolster the case for cosmological
inflation, this set of questions becomes more pressing.  A
complete description of eternal inflation \etinf\ probably
requires an understanding of the many ``universes" behind the de
Sitter horizon of a single inflating patch \refs{\SusskindKW, \dks, \gks }.

Despite the enormous progress in our understanding of quantum
gravity in recent years, physics behind the horizon remains
extremely puzzling.   This is due in part to the holographic
\refs{\StephensAN , \SusskindVU} nature of our most fully
developed nonperturbative descriptions of quantum gravity, the
brane bulk correspondences of Matrix Theory \BanksVH\ and AdS/CFT
\MaldacenaRE, \refs{\GubserBC ,\WittenQJ ,\AharonyTI}. These
formulations describe physics from the point of view of an
observer outside the horizon. For example, CFT correlators in
AdS/CFT describe observables  on the asymptotic spatial boundary
of AdS space. In the thermal CFT that describes a large black hole
in AdS/CFT \WittenZW, these asymptotic observables apparently
describe physics only outside the horizon.

The brane bulk correspondences provide a fully self consistent
quantum description of gravitational dynamics outside of the
horizon, demonstrating that quantum coherence is not lost.  But
these descriptions appear to have little to say about behind the
horizon phenomena.

However, this is not the whole story. A number of authors have
explored ways to extract behind the horizon information from
boundary correlators \refs{\BalasubramanianZV, \LoukoTP,
\MaldacenaKR,  \KrausIV, \LeviCX}.\foot{ There is also large body of work
addressing more general questions of extracting behind the horizon
physics and holographic representation of horizons, as well as of
generic local objects  in the bulk; see e.g.\
\refs{\BanksDD,\BalasubramanianDE,\KabatYQ,\JacobsonMI,\SusskindEY,
\DanielssonFA,\HorowitzFM,\GregoryAN,\GiddingsPT,\FreivogelEX,\HubenyDG}.}
In particular Maldacena \MaldacenaKR\ argued that the boundary
description of the eternal Anti-de Sitter Schwarzschild black hole
 consisting of one copy of the CFT on each of the two asymptotic
boundaries could give some information about physics behind the
horizon. In this approach, the one boundary thermal description is
recovered by tracing over the Hilbert space of the other boundary
CFT as in the real-time or thermofield formalism for thermal field
theory \realtime. The thermal state counting entropy arises from
the entanglement entropy of the pure entangled ``Hartle-Hawking"
state in which expectation values are computed. This realizes an
old idea of Israel \IsraelUR.

These ideas were explored in some detail for $d=3$, the BTZ black
hole, in \KrausIV. The finiteness of amplitudes (manifest from the
boundary or bulk Euclidean point of view) was shown to arise in
the bulk from $i \epsilon$ regulation of the metric singularity
inherent in analytic continuation from the Euclidean theory and
from cancellations between future and past singularities.

In addition, \KrausIV\ gave a simple demonstration of how boundary
correlation functions can  probe physics behind the horizon by
studying the correlator of two operators, one on each asymptotic
boundary, each creating a large mass bulk particle.
As the mass $m \rightarrow \infty$, the correlator can be
evaluated in the semiclassical geodesic approximation and is given
by $\exp(-m \len)$.  Here $\len$ is the proper length of the
spacelike geodesic  joining the boundary points (see Fig.5a). Because
the geodesic passes through spacetime regions inside
 the horizon, this boundary correlator
reveals information about the geometry behind the horizon.  This
closely parallels a prior calculation done in the ``geon" geometry
\LoukoTP.

As for all such quantities, an ``outside the horizon"
interpretation
 can be given.  By tracing the  Hartle-Hawking state, perturbed
by the operator on one boundary, over the Hilbert space of the CFT
on that boundary, a perturbed density matrix for the remaining CFT
is created. This density matrix corresponds in the bulk to a
modified boundary condition on the horizon, which can have
nontrivial correlations with an asymptotic boundary field.  The
nature of this boundary condition is obscure from the ``outside
the horizon" point of view, but natural when physics behind the
horizon is taken into account.

The BTZ geodesics illustrated in Fig.5a pass quite close to the
singularity for moderate boundary times.  But the correlation
function is relatively structureless.  Presumably this reflects
the simple orbifold geometry of the BTZ black hole.  The curvature
of the BTZ geometry is constant except for a delta function at the
singularity.

Our goal in this paper is to explore further what information from
behind the horizon can be found in boundary AdS/CFT correlators.
To consider a situation with a more interesting  geometry, we
focus on AdS Schwarzschild black holes  in $d>3$ , and in
particular in $d=5$ where the boundary CFT is four dimensional
${\cal N }=4$ super Yang Mills theory.  The geometry near the
singularity in such black holes approaches that of $d>3$
Schwarzschild, and so the curvature diverges as one approaches the
singularity. This affects the geodesics dramatically.

We begin in Section 2 by exploring in detail the causal structure
and spacelike geodesics in $d=5$ \sads.  (Other dimensions
greater than three behave similarly.) We find several unusual
features. First, the Penrose diagram that encodes the causal
structure of the geometry is not a square, unlike the $d=3$ case.
The lines describing the singularity bend inward toward the center
of the diagram.  This allows a nearly null geodesic beginning at
the boundary at a finite nonzero time $t_c$ to ``bounce" off the
singularity and hit the other boundary at the symmetric point.
This behavior, which does not occur for $d=3$,  is depicted in
Fig.5b.

After submitting this paper we have learned that such features have previously
been found in a general study of effectively two dimensional metrics \KloschQV\ \foot{We are grateful
to Jorma Louko for bringing this reference to our attention.} .

In Section 3 we turn to correlators in the geodesic approximation.
If the bounce geodesic were to dominate, the correlator  would
become singular as $t \rightarrow t_c$ because the proper distance
 goes to zero as the geodesic becomes null. This would be a kind
of ``light cone singularity." But general considerations about the
boundary field theory rule out this kind of behavior. In fact this
geodesic does {\it not} dominate the correlator. There are in
general multiple geodesics that connect the two boundary points (Fig.12).
At $t=0$ their proper distances coincide, creating a branch point
in the correlator which behaves as $t^{4/3}$ for small $t$. By
studying various resolutions of this branch point, we show that as
$t$ increases from $0$, the correlator defined by the boundary CFT
is given by a symmetric sum of the two complex branches of this
expression. Each of these can be attributed to a complex geodesic
in complexified spacetime. At large $t$ these complex branches
reproduce the correct quasinormal modes for this black hole, which
are complex, unlike the $d=3$ case.

But the correlator is an analytic function of $t$ and can be
continued onto the real sheet.  (This can be done  in a
computationally effective manner, as discussed in Appendix F.) On
this sheet, the ``light cone singularity"   does appear. So the
boundary correlator {\it does} contain information about the
singularity, albeit in a subtle way.

In Section 4 we extend the above analysis to include finite $m$,
finite string length $l_s$ ($\alpha '\sim l_s ^2$), and finite
string coupling ($g_s$) effects. We take wrapped D-branes (giant
gravitons \McGreevyCW)  as explicit examples of heavy probe
particles. The branch point at $t=0$ is smoothed out for any
finite $m$, preventing analytic continuation to the bouncing
geodesic. But at each order in the $1/m$ expansion the branch
point persists and one can follow this geodesic and the
accompanying fluctuation corrections into the bouncing domain. The
$1/m$ corrections are given by a heat kernel expansion with
coefficients related to the curvature and are large when the
geodesic passes near the singularity. These quantities give a
clear example of information about the singular geometry being
coded into boundary CFT correlators.

The main corrections at finite but small $l_s$ (with $g_s$ kept
equal to zero)  can be expressed as modifications of the
supergravity field equations and hence of the background metric,
dilaton, etc.\ fields.   Such small modifications do not change the
basic picture outlined above.  The branch point at $t=0$ either
persists or, more generically, splits into two nearby branch
points.  In either case we can follow the geodesic onto the real
sheet.  Of course, for $t$ near $t_c$, we expect the $l_s$
corrections to become large because the geodesic passes through
regions of large curvature.

At large $l_s$ the 't Hooft coupling $\lambda$ in the boundary
gauge theory becomes small.   In weak coupling perturbative gauge
theory amplitudes are smooth at $t=0$.  There is no branch point.
Therefore, for consistency, there must be a phase
transition at finite $\lambda$ (for infinite $m$).

Finite $g_s$ is a particularly difficult regime in which to study
these phenomena. Wrapped D-brane masses are $\sim 1/g_s$ and so
are finite here.  This smooths out the branch point and would seem
to prevent analytic continuation to the $t_c$ singularity. But by
taking appropriate $g_s \rightarrow 0 $ limits we can extract
behavior around the singularity to all orders in $g_s$, as well as
study certain leading nonperturbative effects.

These results demonstrate that a significant amount of information
from behind the horizon, and in particular from near the
singularity, is encoded in  boundary theory correlators.  In
Section 5 we discuss the meaning of these results further.

\newsec{Classical geometry}

We will devote this section to the exploration of the classical
Schwarzschild-AdS geometry. As is well known, the \sads\
spacetimes contain spacelike curvature singularities, event
horizons, and timelike boundaries, all of which will play an
important role in the discussion.  First, we analyze the causal
structure in detail, arriving at the surprising observation that,
unlike the three dimensional case, for $d>3$ the Penrose diagram
of \sads\ spacetimes cannot be drawn as a square, with both
boundaries and singularities represented by straight lines. Next,
we study  spacelike geodesics in this geometry, which, as we will
discuss in the next section, provide a method for approximating
the relevant boundary correlators.

\sads\ represents a $d \ge 3$ dimensional, two-parameter family of
solutions characterized by the size of the black hole and the AdS
radius.  For simplicity we will focus on the large black hole
(planar) limit, where 
the  AdS radius is much smaller than the horizon radius.
Also, we will from now on concentrate on $d=5$.  The geometry
for finite-sized black holes, as well as in other
dimensions $d>3$, is qualitatively similar.
We will contrast this with the quite
different case of the $d=3$ BTZ black hole \KrausIV .

The metric of the 5-dimensional \sads\ spacetime is given by

\eqn\metric{\eqalign{ ds^2 &= - f(r) \, dt^2 + {dr^2 \over f(r)} +
r^2 \, d\Om_3^2 \cr f(r) &= {r^2 \over R^2} + 1 - {\rh^2 \over
r^2} \( {\rh^2 \over R^2} + 1 \),}} where $\rh$ is the horizon
radius and $R$ is the AdS radius.  We work in the limit of the
infinitely massive black hole $r_+/R \rightarrow \infty$, where
the metric simplifies. In particular, if we rescale the
coordinates $r \rightarrow {\rh \over R} \, r, ~ t \rightarrow {R
\over \rh} \, t$,  measure lengths in AdS units so $R=1$,  and
suppress the angular coordinates $\Omega$ that will not concern us
here, the metric in the $t$-$r$ plane becomes

\eqn\metric{\eqalign{ ds^2 &= - f(r) \, dt^2 + {dr^2 \over
f(r)} \,  \cr f(r) &= r^2 - {1 \over r^2}\, .}}

There is a genuine curvature singularity at $r=0$, and the
boundary of the spacetime is approached as $r \to \infty$. The
Schwarzschild coordinates used in \metric\ also have a coordinate
singularity at the horizon $r=1$. This is not a big obstacle,
since one can pass to Kruskal  coordinates $(T, X)$ which cover
the full globally extended spacetime.  (The Kruskal coordinate
chart is constructed explicitly for this geometry in Appendix A).
However, the global extension can be discussed more conveniently
for our purposes by using four Schwarzschild coordinate patches
(corresponding to the two asymptotic regions (I and III) with
$r>1$, plus the regions inside the black hole (II) and the white
hole (IV) with $0<r<1$; cf.\ Fig.1).

\ifig\figSchcoords{Complexified coordinates for the AdS black
hole: the time coordinate is complex in the extended spacetime,
but with constant imaginary part in each wedge, as indicated
above.  The wedges are separated by the horizons at $r=1$, where
$t$ diverges.  Note that the real part of $t$ increases upward in
wedge I, downward in wedge III, and to the right (left) in wedge
II (IV).} {\epsfxsize=7cm \epsfysize=5.4cm \epsfbox{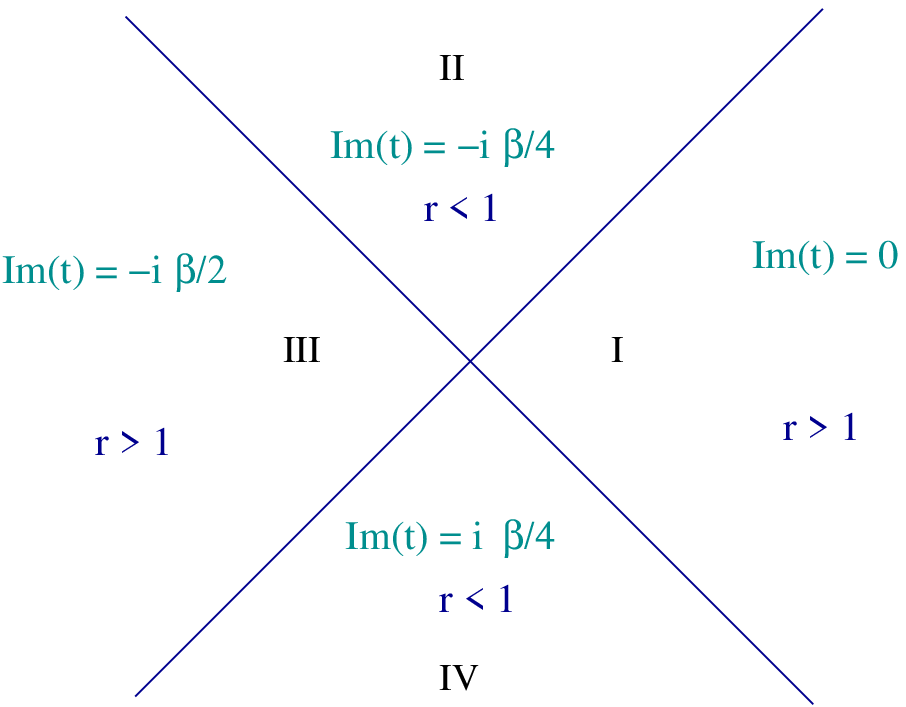}}

These four patches can be embedded in complexified
Schwarzschild time, \eqn\compltime{t= t_L +i \, t_E}
 where $t_L$ and $t_E$ denote times on
the Lorentzian and Euclidean slices respectively. In each
coordinate patch, $t$ has a constant imaginary part, as indicated
in \figSchcoords; we will define this to be 0 for the right
asymptotic region. Crossing a horizon shifts the imaginary part by
$i \, \beta/4$.  Setting $t=i \, t_E $ produces Euclidean \sads\
with a periodic imaginary time coordinate $t_E$. The period of
$t_E$ is $\beta$, the inverse temperature $T$ of the black hole.
One can move a point from the boundary of region I to the
symmetric point on the boundary of region  III in \figSchcoords\
by rotating by half a period; that is, shifting $t \rightarrow -t
- i \, \beta /2$. The minus sign accounts for the opposite
direction of time in region III. In the geometry \metric ,
$\beta=\pi$. The CFT correlators can be defined for complex time
and are analytic, so such an extension is natural from the
boundary side. The periodicity is just that of finite temperature
field theory.


\subsec{Radial null geodesics}

We now turn to a major focus of this paper, the study of geodesics
in the geometry \metric .  We consider a radial ($\Omega$
constant) path,  $x = (t(s), r(s))$, where $s$ is an affine
parameter. Geodesic paths can be found by extremizing the action
\eqn\geoact{S = \int ds ~{\dot x}^2,} where ${\dot x = {d x \over
d s}}$. We can then find a unit speed parametrization by requiring
${\dot x}^2 =0, +1, -1$ for null, spacelike, and timelike
geodesics, respectively.

Due to the symmetry described by the Killing field $\dt$, there
exists a conserved quantity $E$ along each geodesic, \eqn\Ecm{ E =
\td \, f(r). } To map out causal structure we first consider null
geodesics. In a spherically symmetric spacetime, radial null rays
will always be 45-degree lines in the Penrose diagram. Here ${\dot
x} ^ 2 = -f (r) \, {\td}^2 + {{\rd}^2 \over f(r)} = 0$, which when
combined with \Ecm\ becomes \eqn\rdotsq{ \rd^2 = E^2~ .}

Let us consider the coordinate time $t(r)$ along an ingoing radial
null geodesic which starts from the boundary $r = \infty$ at
$t=t_0$. From \Ecm\ and \rdotsq\ we have that

\eqn\tra{ t(r)  = \int {\td \over \rd'} \, dr' = t_0 +
\int^{\infty}_{r} {dr' \over f(r')}.}
 For geodesics which cross the
horizon, integrating over the pole at $r = 1$ gives a constant
imaginary part $-i {\pi \over 4}$, so that for $r < 1$, \eqn\tre{
t(r) = t_0 - {1 \over 2} \[ \tan^{-1} r - {\pi \over 2} -
\tanh^{-1} r
\] - i \, {\pi \over 4}. } The the last term is simply $i \beta
/4$, where again $\beta = 1/T = \pi$ is the Bekenstein-Hawking
temperature (cf.\ \figSchcoords) for this geometry.

This result implies that a null geodesic which starts at the
boundary at $t_0 = 0$ reaches the singularity $r=0$ at
$ 
 t(0) = {\pi \over 4} \( 1 - i \)$. 
Since the real part of $t(0)$ is positive,
the geodesic hits the singularity {\it off-center}
in the Penrose diagram; in fact it hits closer to the right
boundary, where it originated.

\ifig\figPdg{Behaviour of radial null geodesics suggests the shape
of the Penrose diagram (wavy lines represent the singularities,
dashed lines the horizons, vertical lines/curves the boundaries,
and the solid 45 degree lines the radial null geodesics): a) if
geodesics meet at the singularity, the diagram can be drawn as a
square; if they don't meet, b) with straight boundaries, the
singularities are bowed in, or c) with straight singularities, the
boundaries must be bowed out.} {\epsfxsize=12cm \epsfysize=3.7cm
\epsfbox{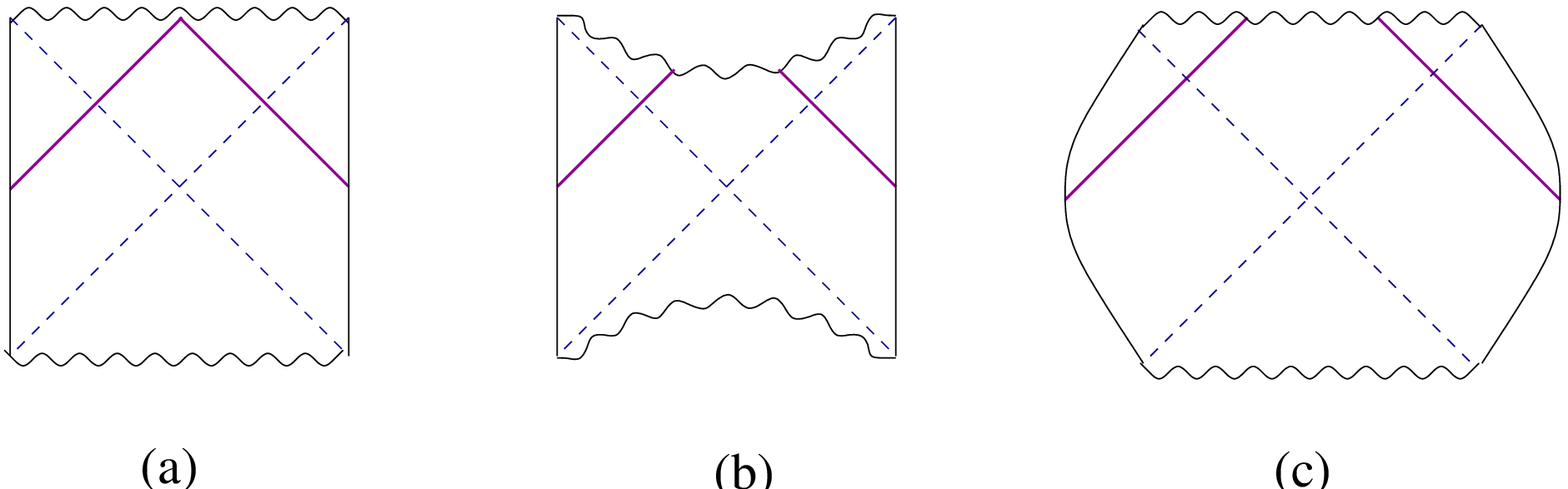}}

This immediately suggests the nature of the Penrose diagram, as
illustrated in \figPdg. If the Penrose diagram were a square, as
in \figPdg a, then radial null geodesics sent off from the
boundary at $t=0$ (defined as the horizontal line of symmetry of
the Penrose diagram) would have to meet at the same point on the
singularity. On the other hand, if the geodesics don't meet at or
before reaching the singularity, then the
singularity on the Penrose diagram must be drawn {\it bowed
in}, assuming we draw the boundaries straight, as in \figPdg b.
 Alternately, if the singularity were to be
represented by a straight line, the boundaries would have to be
bowed out,  as in \figPdg c, in order to reproduce the calculated
behaviour of the geodesics. Note that the Penrose diagrams
sketched  in \figPdg b and \figPdg c are equivalent, since we can
use a conformal transformation to straighten out the singularities
of \figPdg b, but at the expense of bowing out the boundaries.
This will be discussed in greater detail in Appendix B, where the
Penrose diagram for this spacetime is constructed explicitly.
Such Penrose diagrams have previously been discussed in \KloschQV\ .

Conversely, we can ask at what time does a null geodesic have to
start from the boundary in order to reach the singularity at
$\Re[t(r=0)] = 0$, i.e.\ in the ``middle'' of the singularity.
This will play an important role later. Denoting such a time by
$\tc$, we see that \eqn\tcrit{ \tc = - {\pi \over 4} = - {\beta
\over 4} .} In general dimension $d$, one finds that $t_c = - \pi
 \( (d-1) \, \tan {\pi \over (d-1)} \)^{-1}$, which vanishes only for
$d=3$. Thus we see that, with the exception of the $d=3$ BTZ black
hole, all singularities in the Penrose diagrams of \sads\ are
bowed in as in \figPdg b.

\ifig\krusk{Radial null geodesics on Kruskal diagram, a) for
$d=3$, and b) for $d>3$ (radial null lines lie at 45 degrees, the
horizons are represented by dashed lines, and the boundaries and
singularities by the solid hyperbolas).  In order for radial null
geodesics to meet at the singularity at $X=0$, they have to start
at a) $t=0$ for the $d=3$ case, and b) $t = - t_c < 0$ for the
$d>3$ case. } {\epsfxsize=12cm \epsfysize=6.3cm
\epsfbox{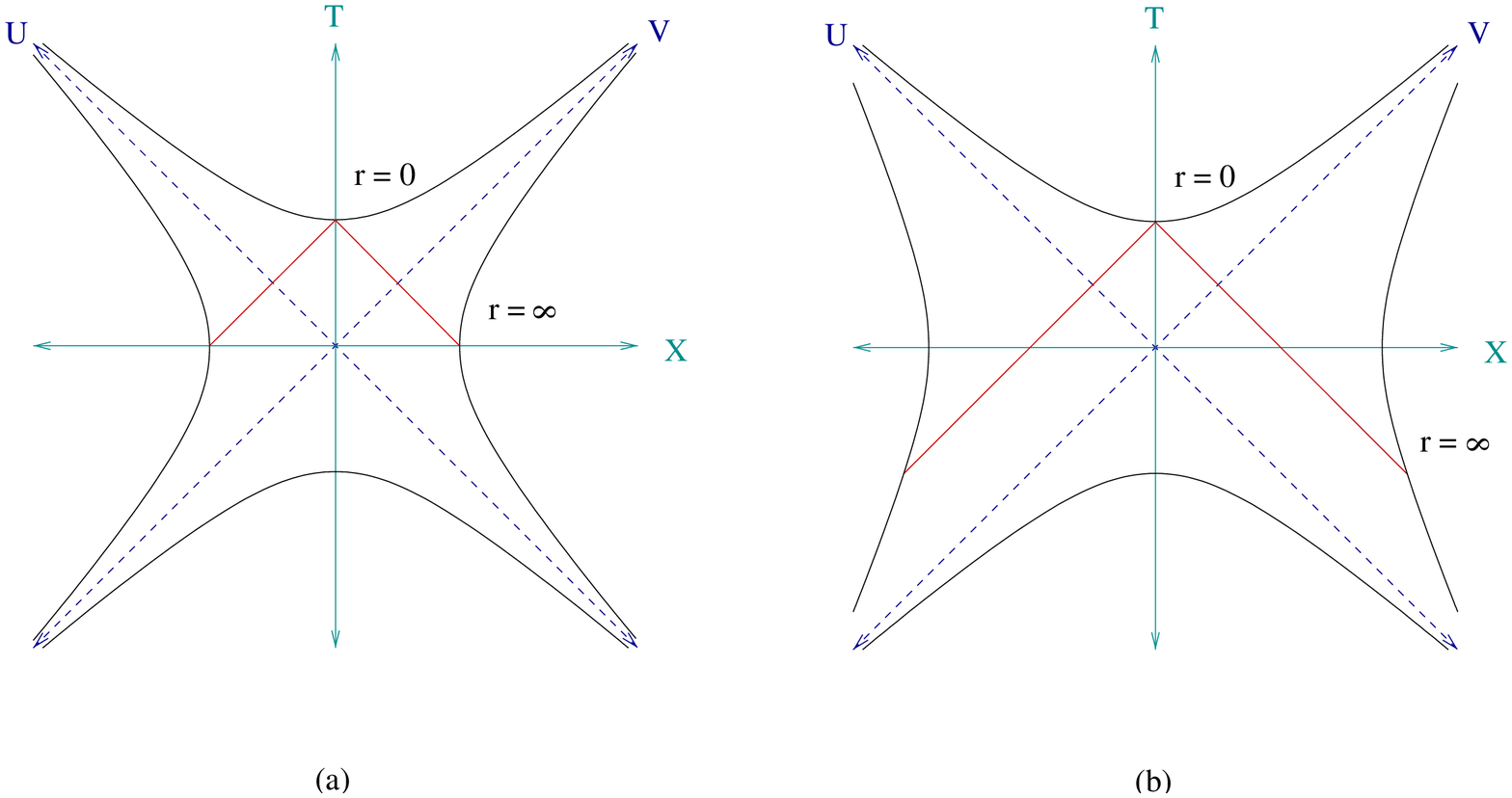}}

In Kruskal coordinates $(T, X)$, the future and past singularities
of the black hole are the hyperbolas $T^2 - X^2 = 1$, while the
AdS boundaries are given by $T^2 - X^2 = - e^\pi$. (By contrast,
in the $d=3$ BTZ geometry, the singularities are at $T^2 - X^2 =
1$ and the boundaries are at $T^2 - X^2 = - 1$.)  The curved nature of
the boundary of the Penrose diagram is reflected in these
coordinates by the asymmetry in the radii of the two hyperbolas.
This asymmetry allows a null geodesic starting at $t_c$ to
``bounce off" the singularity.  This is illustrated in \krusk ,
where the $d=3$ case (\krusk a) is contrasted with the
higher-dimensional case (\krusk b).
In Appendices A  and B we explain this in further detail, and construct
the Penrose diagram explicitly.  (The latter will be used in the
next subsection to produce the Penrose diagrams in Fig.5.)
These issues are also discussed in \KloschQV\ .

\subsec{Radial spacelike geodesics}



Let us now turn from the qualitative structure of the Penrose
diagram to quantities which we can probe from the CFT side: in
particular, the proper length along spacelike geodesics. In this
subsection, we will consider purely radial spacelike
geodesics.\foot{For completeness, we
 present in Appendix C  an analogous treatment
 for spacelike geodesics carrying angular momentum.  For the
 remainder of the paper, however, we need only the results about
 geodesics with zero angular momentum.}
 The spacelike geodesics satisfy ${\dot x} ^ 2 = -f (r) {\td}^2 + {{\rd}^2 \over
 f(r)} = 1$, giving \eqn\spgeod{ E = \td \, f(r), \ \ \ \ \ \ \rd^2 -f(r)
= E^2 .}  Here $E$, as before, is a conserved quantity resulting
from the time translation symmetry of the geometry. We can think
of it as the total energy of a particle moving in a potential
$V(r) = -f(r)$ (where the minus sign is due to the fact that we
are considering spacelike geodesics).

\ifig\energy{The ``potential energy" in the mechanics problem for
spacelike geodesics in \sads\ in $d=3$ and $d=5$. The potential
diverges at $r=0$ for all $d > 3$.} {\epsfxsize=7cm
\epsfysize=5.9cm \epsfbox{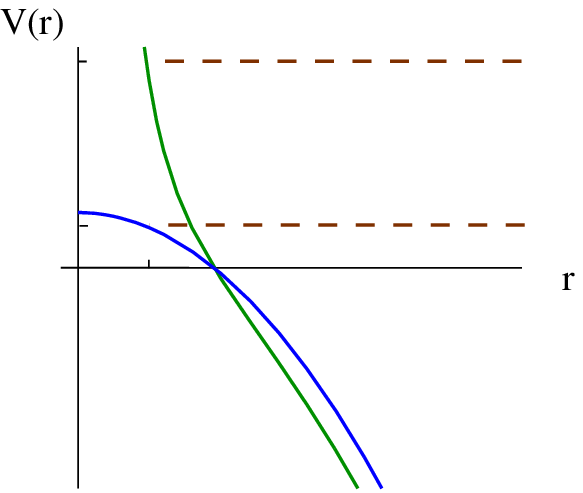}}

In \energy, we plot the potential $V(r)$ for $d=3$ and $d=5$
\sads. In the BTZ case, the energy required to reach $r=0$ is
finite. As a result, the proper time along the geodesic is large.
By contrast, in the higher dimensional examples the potential
diverges at the singularity, so that  large $E$ is required to
approach $r=0$.  At large $E$ the particle is moving fast and so
very little proper time is covered. The result as $E \rightarrow
\infty $ is a geodesic that is null everywhere except in a
vanishingly small region very near the singularity where it
``bounces" off. As explained above, this occurs at $t_0 = t_c$.

\ifig\figgeods{Symmetric spacelike radial geodesics of the \sads\
black hole, plotted on the corresponding Penrose diagrams. a)
$d=3$: Because only finite $E$ is required to come arbitrarily
close to the singularity, none of these geodesics is nearly null.
b) $d>3$: the spacelike geodesics converge to a large $E$ almost
null geodesic which comes arbitrarily close to the singularity. }
{\epsfxsize=14cm \epsfysize=7.5cm \epsfbox{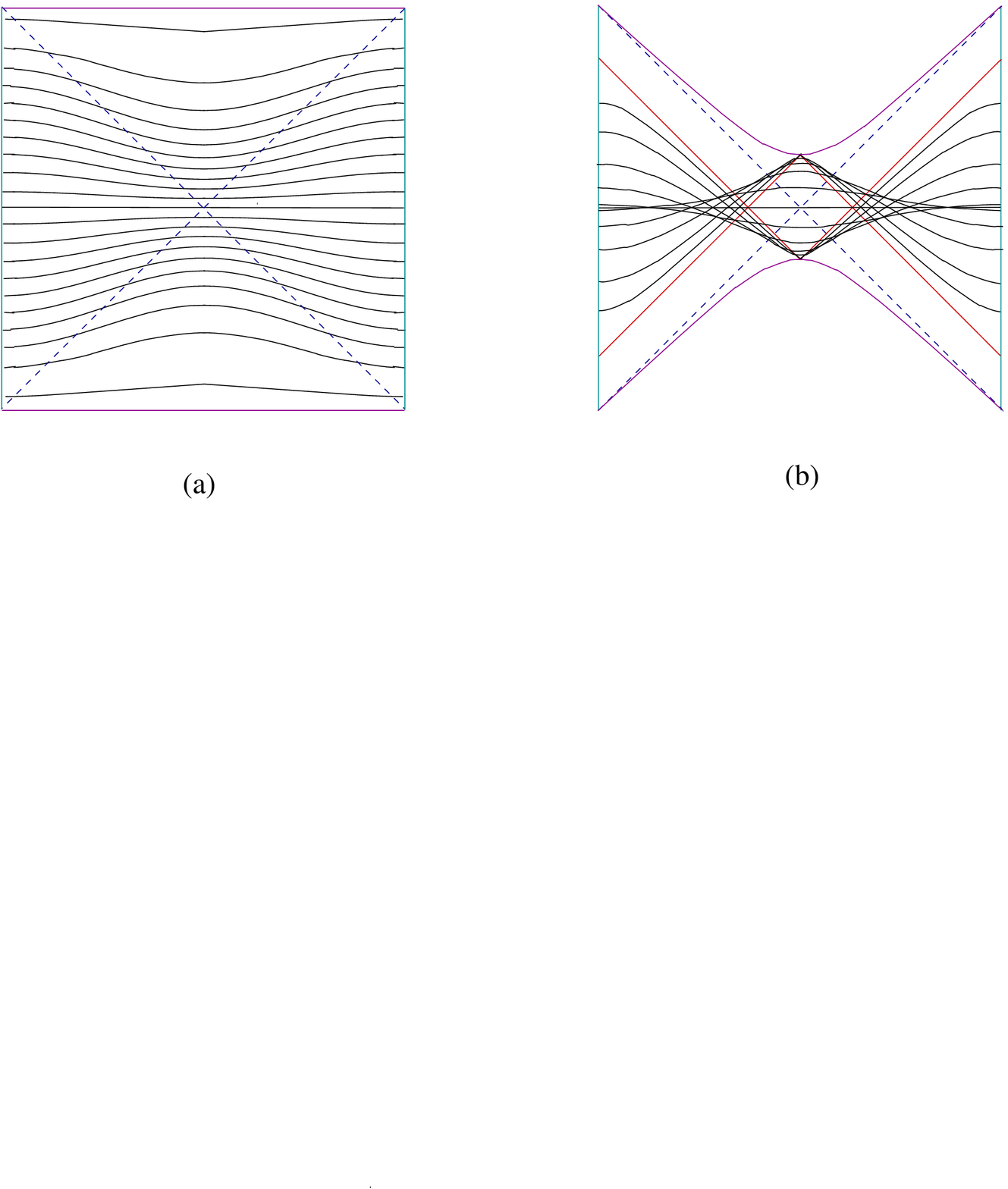}}

The behaviour of radial spacelike symmetric geodesics (i.e. those
which reach their minimum value of $r$ at $t= {i \beta \over 4}$)
 in $d=3$ and $d=5$ is plotted on the corresponding Penrose
diagrams in \figgeods. As explained above, in the $d=3$ BTZ case
(\figgeods a), the geodesics do not show any striking feature as
the starting time $t_0$ is varied.  On the other hand, in higher
dimensions such as $d=5$ (\figgeods b), spacelike symmetric
geodesics exist only for a finite range of starting times, $t_0
\in (-t_c, t_c)$. For starting times outside this range, no such
geodesics exist. This is in accord with the expectation that the
genuine curvature singularity of the higher dimensional black
holes has a pronounced effect on geodesics in its vicinity, as
opposed to the $d=3$ BTZ orbifold singularity.

Let us first consider the $E=0$ case. From \spgeod, we see that
$\td = 0$; $t(s) = 0$ is the only consistent solution.  We have
that $\rd^2 = f(r)$. Such a geodesic crosses the Penrose diagram
in the middle, so that it goes from $r=\infty$ at one boundary,
down to the horizon $r = 1$, and back out to $r=\infty$ at the
other boundary.  Since it never penetrates inside the horizon,
this geodesic can equally well be embedded in Euclidean \sads.
It is the geodesic that crosses over the tip of the cigar to the
antipodal point $t_E = \beta/2$ on the Euclidean time circle (Fig.
8 ).

The proper length along such a geodesic is then given by
\eqn\length{ \len = 2 \int_{1}^{\rinf} {dr \over \rd} = 2
\int_{1}^{\rinf} {dr \over \sqrt{f(r)} }} where $\rinf \to \infty$
is the large $r$ regulator standard in AdS/CFT \AharonyTI\ .
Integrating this explicitly, we have \eqn\Lresc{
\len  
= \ln \( \rinf^{2} + \sqrt{ \rinf^4 -1} \) \approx 2 \ln \rinf +
\ln 2 + \CO(\rinf^{-4}) } so that taking $\rinf \to \infty$ and
subtracting off the universal logarithmically divergent piece, we
obtain the regularized proper length \eqn\Lreg{ \len_{\rm reg} = \ln
2 } In the future we will drop the subscript on $\lenr$.

Let us now consider what happens as we increase the energy $E$.
From \spgeod\ we see that the geodesic now penetrates a finite
distance inside the horizon. Specifically, let $\ri$ denote the
smallest value of the radial coordinate reached along such a geodesic. This
corresponds to the
classical turning point of the particle motion, which is
obtained by solving \eqn\rieqn{ E^2 + f(\ri) = 0 ~.} Since $f(r)$
is positive outside the horizon and negative inside, $\ri \le 1$.
For very high energy $E \gg 1$, the geodesic comes very close to
the singularity, $\ri \ll 1$, so that $f(r) \sim - {1 \over r^2}$,
which means that $\ri \sim {1 \over E}$.

In general, spacelike geodesics with arbitrary $E$ and starting
time $t_0$ will not be symmetric.  We can specialize to this case,
however, and obtain all others by $t$ translation.  For a given
$E$, there exists a starting time $t_0$ on the boundary in region
I  for which the geodesic is symmetric. In particular, for a
symmetric geodesic the turning point $r = \ri$ has to occur at
$\Re[t(\ri)] = 0$. The turning point must be in region II so
$\Im[t(\ri)]= -\beta/4$. The geodesic ends in region III with
coordinate $t= -t_0 - i \beta/2$. As mentioned before, the real
part has a minus sign because the Schwarzschild coordinate time
runs in the opposite directions in regions I and III. The geodesic
equations give \eqn\tstartsp{ t_0-t(\ri)=t_0+ i \beta/4 =  -
\int_{\ri}^{\infty} {\td \over \rd} \, dr
 =   - \int_{\ri}^{\infty} {E \over f(r) \, \sqrt{ E^2 +
f(r) }} \, dr. } For $E=0$, we recover $t_0=0$. One can also check
that as $E \to \infty$, $t_0 \to \tc$; and in fact, for large
energies one can see from examining the integral that \eqn\tdiff{
(t_0 - \tc) \sim { 1 \over E }.} For  each such geodesic there
exists a mirror one obtained by taking $t \rightarrow -t, ~ E
\rightarrow -E$.

The proper length, analogous to \length, along a symmetric
spacelike geodesic originating at $\rinf$, going down to $\ri$,
and then back out to $\rinf$ in the other asymptotic region, is
\eqn\lengthE{ \len = 2 \int_{\ri}^{\rinf} {dr \over \sqrt{E^2 +
f(r)} }\ .}  Upon evaluating and regularizing this integral, we
obtain \eqn\lengthEr{\len = \ln \({2 \over \sqrt{{E^4} / 4 + 1}}\)
.}  As $E$ goes to infinity the geodesic looks more and more null,
and so at any fixed radial cutoff its proper length goes to zero.
This is apparent already from \figgeods b, where we can see
explicitly that symmetric
spacelike geodesics in the $E \to \infty$ limit approximate the
null geodesic which bounces off the singularity.
The regularized length along an $E \to \infty$ geodesic
 (which is roughly the difference between
its proper length and that of the $E = 0$ geodesic) goes to
negative infinity at $t_c$,

\eqn\lengthErapp{\len \sim 2 \ln \(t-t_c\).}
We will make use of this result in the next section, after we discuss
the correlation functions.


\newsec{Correlation functions}

According to the AdS/CFT correspondence, \sads\ in $d=5$ is dual
to
${\cal N}=4$ SYM theory at finite temperature.  
As before, we will consider the large black hole limit, so the CFT
effectively lives in infinite volume. Standard CFT finite
temperature boundary correlators are dual to bulk correlation
functions with insertions all in one asymptotic region (e.g.,
region I).    There are other boundary correlators one might want
to study.   For example, one could put an operator on the boundary
of region I, and another on the boundary of region III
\refs{\MaldacenaKR , \KrausIV} .  We have discussed the geodesics
connecting such points.  In the field theory, such correlators
have a natural representation in the thermofield double or
real-time description of finite temperature field theory
\realtime.

In the real-time formalism, one takes the tensor product of two
copies of the original field theory labeled by $1,2$.  The two
copies are decoupled, and the total Hamiltonian is \eqn\Hdiff{
H_{tf} \equiv H \otimes I - I \otimes H^*, } where $H$ is the
Hamiltonian for the original theory.  We now construct the
entangled state \eqn\state{ |\psi \rangle= {1 \over Z^{\12}}
\sum_i e^{-\12\beta E_i}|E_i,E_i\rangle, } where
$|E_i,E_j\rangle=|E_i\rangle\otimes|E_j\rangle$, and $|E_i\rangle$
are energy eigenstates. The state $|\psi\rangle$ is a particular
eigenvector of $H_{tf}$ with eigenvalue zero.   Correlations
between subsystems $1$ and $2$ are due to the entanglement in
$|\psi \rangle$.

Operators which belong to subsystem $1$ have the form $A \otimes
I$ (where $I$ is the identity operator), and will be denoted
$A_1$. Operators associated with subsystem $2$ are defined in a
similar manner, except with an additional rule of hermitian
conjugation: \eqn\atwo{ A_2 \equiv I \otimes A^{\dag}. } Standard
thermal correlation functions may be written as an expectation
value: \eqn\tfcora{ \langle \psi | A_1(0) B_1(t) | \psi \rangle .}
As can be easily seen from the form of $ | \psi \rangle$, \tfcora\
is simply the thermal expectation value of $A(0) B(t)$, evaluated
in a thermal density matrix at inverse temperature $\beta$.  The
state counting entropy observed in subsystem $1$ is the entropy of
entanglement of the state $|\psi \rangle$.  In field theory, no
physical significance is usually attached to correlators involving
both subsystems, but we can certainly define them; for example

 \eqn\tfcorb{ \langle \psi| A_1(0) B_2(t) |\psi\rangle. }
In the finite temperature AdS/CFT correspondence, \tfcorb\ has a
simple interpretation \MaldacenaKR : it corresponds in the bulk to
a correlator  between operators on the two disconnected boundaries
of the spacetime.

It is not hard to see that one can compute \tfcorb\ by
analytically continuing \tfcora.

\eqn\tfcorc{ \langle \psi| A_1(0) B_2(t) |\psi\rangle = \langle
\psi | A_1(0) B_1(-t-i\beta/2) | \psi \rangle . }
This is the analog of using complexified Schwarzschild time as
discussed in Section 2.   We see that these two sided correlators
are just part of the information contained in ordinary thermal
correlators, as a function of complex time.

We can now use our knowledge of geodesics in \sads\ to study
correlation functions in the CFT.  The prescription is to
determine the 2-point CFT correlators via AdS/CFT from a
computation of the bulk propagator.  We assume at this point that
we have a scalar field of mass $m$ in the bulk, dual to some
operator in the CFT whose 2-point function we want to compute. The
bulk propagator is given by a sum over paths between two points in
the bulk, with each path contributing $e^{-m \len}$, $\len$ being
the proper length of the path, suitably regulated.  In the limit
of large $m$, this sum will generically be dominated by the
shortest geodesic connecting the two points. So the 2-point CFT
correlator should go like $e^{-m \len}$ (up to $1 \over m$
corrections), where $\len$ is the regularized length of the
shortest geodesic between the two points\foot{This approach and
some of the pitfalls involved are discussed in \LoukoTP.}.

In section 3.1 we show that a naive application of this
prescription predicts a ``light cone'' pole in the 2-point
opposite side correlator  as $ t \to t_c$.   However, simple
arguments in the field theory show that this prediction cannot be
correct, at least in the theory defined by analytic continuation
from Euclidean space. In Section 3.2 we address this puzzle by
examining the branch cut singularity in $\len$ at $t=0$ that is
due to a coalescence of geodesics.  We first resolve the branch
cut by taking $\rinf $ large but finite. We find that the
unambiguously correct geodesic in Euclidean space bifurcates into
two complex geodesics at small Lorentzian time $t_L$.  These
geodesics must determine the correct answer.   In section 3.3 we
elucidate this further by moving this bifurcation onto the
Euclidean section. This involves taking the black hole mass large
but finite. Euclidean correlators are unambiguously determined by
the shortest geodesics at large $m$, and so we can determine the
correct choice of geodesics through the bifurcation. The dominant
geodesics continue to the complex ones mentioned above as one goes
to the Lorentzian section. However, at large $m$ it is possible to
analytically continue correlators around these bifurcations by
following geodesics that are no longer dominant. This allows us to
find the $t_c$ singularity on a secondary sheet of the branched
correlation function.  This $t_c$ singularity gives information
about the black hole singularity.  It shows that such information
is encoded in CFT correlators, albeit in a subtle way involving
analytic continuation.  In section 3.4 we introduce a simple model
of the bifurcation given by an ordinary integral. The properties
of this model can be determined precisely and agree with the
picture we have obtained. We demonstrate here that all orders
fluctuation corrections around a given saddle point can be
analytically continued as well.  This indicates that additional
information about the black hole singularity is also accessible.
In section 3.6 we show that our results for the correlator agree
with the known values for the quasinormal mode frequencies of the
\sads\ black hole.  This gives independent confirmation of our
analysis.

\subsec{A light cone singularity }

Recall from Equation \lengthErapp\ that the regularized length of
a symmetric geodesic with boundary time $t$ diverges
logarithmically at $t_c$. Therefore the correlation function
behaves like \eqn\lcsing{e^{-m \len} \sim {1 \over (t -
t_c)^{2m}}\ .} We seem to have a pole of order $2m$ in the
two-sided
CFT correlator coming from the ``almost null'' 
geodesic connecting the boundary points.

This immediately leads to a puzzle.  We know that the real-time
two-point correlation function, evaluated in a thermal state with
inverse temperature $\beta$, is \eqn\twopoint{\eqalign{ \langle
\phi(t) \phi(-t) \rangle_\beta  = \sum_n e^{- \beta E_n} \langle
E_n | e^{- i H t} \phi(0) e^{ i H t} e^{ i H t} \phi(0) e^{- i H
t}| E_n \rangle \cr = \sum_{n,m} e^{ -\beta E_n - 2 i t (E_n -
E_m)}
 \left| \phi_{nm}   \right|^2 ,}}
where $\phi_{nm} \equiv \langle E_n | \phi(0) | E_m \rangle$ and
we have assumed $\phi^{\dag} = \phi.$  To get the two sided
result, we can either use the real-time formalism and the state $|
\psi \rangle$, or continue one of the times from the real value
$t$ to $t - i \beta /2$: \eqn\twopointagain{  \langle \psi |
\phi_2(t) \phi_1(-t) | \psi \rangle = \langle \phi(t + i \beta/2 )
\phi(-t) \rangle_\beta  = \sum_{n,m} e^{ - {\beta \over 2} (E_n +
E_m) - 2 i t(E_n - E_m)} \left| \phi_{nm}  \right|^2. }  Note that
the terms in the sum are real and positive for $t=0$;  therefore
\eqn\ineq{\left| \langle \phi_2(t) \phi_1(-t) \rangle_\beta
\right| \leq \langle \phi_2(0) \phi_1(0) \rangle_\beta} for any
Hermitian operator $\phi$.  However, $\langle \phi_2(0) \phi_1(0)
\rangle_\beta$ can be computed unambiguously in Euclidean space,
using the geodesic approximation. It is certainly finite (since
the two points are on opposite sides of the Euclidean cigar). Thus
we cannot have a singularity in the opposite side correlator at
any time, including $t = t_c$.

\subsec{Tracking geodesics}

We begin exploring this puzzle by computing the boundary time as a
function of $E$ from the integral \tstartsp .  The computation is
detailed in Appendix D; the result is

 \eqn\longeqn{\eqalign{t &=
{\half} \ln {\({{{\half} E^2 - E + 1} \over {\sqrt{1 + {\quarter}
E^4}}}\)} - {\half} i \ln {\({{- {\half} E^2 + i E + 1} \over
{\sqrt{1 + {\quarter} E^4}}}\) } \cr \len &= \ln \({2 \over
\sqrt{1 + {\quarter} E^4}}\) ~ ,}} where formula \lengthEr\ has been
repeated for convenience. We have not restricted ourselves to
symmetric geodesics here. The time $t$ is the difference between
$t_f$ and $t_i$, the final and initial boundary Schwarzschild
times; $t=t_f-t_i + i \beta/2$. We have chosen the origin so that
$t=0$ describes endpoints in regions I and III at zero time.  For
symmetric geodesics $t=-2 \, t_0$.

These equations determine $\len (t)$ and hence the large $m$
correlator $\langle \phi \phi \rangle (t) = \exp(-m \len(t))$.
They are central to our analysis.

Expanding \longeqn\ around $E=0$ we find \eqn\Eexpans{\eqalign{ t
&\sim E^3 \cr \len &\sim - E^4 \cr \len &\sim -t^{\ft}~.}} The
branch point at $t=0$ signals novel behavior.  There are three
branches for $\len$ and hence for the correlator at each $t$.  It
will turn out the gauge theory chooses a symmetric combination of
the two complex branches. The $t_c$ singularity lies on the real
branch. But knowing the correlator on the complex branch will
allow us to study the real branch by analytic continuation.
Information about the black hole singularity is encoded in the
gauge theory in this manner.  The analytic structure of the
Riemann surface defined by \longeqn\ is discussed in Appendix E.

\ifig\cubic{Plot of $t$ as a functions of $E$, normalized by
$t(E \to \infty)$, for various values of $\rinf$:
the solid line has $\rinf = 2$, while
the dashed line represents $\rinf = \infty$}
{\epsfxsize=7cm \epsfysize=3.2cm \epsfbox{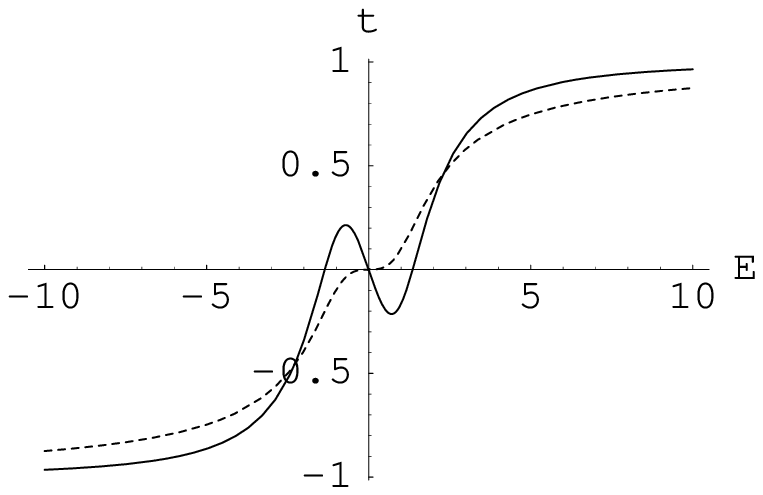}}

\ifig\threegeods{Three ``zero time'' geodesics  between the two
cutoff boundary points ($t=0, r=\rinf$).  The curved dashed lines are
$r = \rinf$.}
{\epsfxsize=6cm \epsfysize=6cm \epsfbox{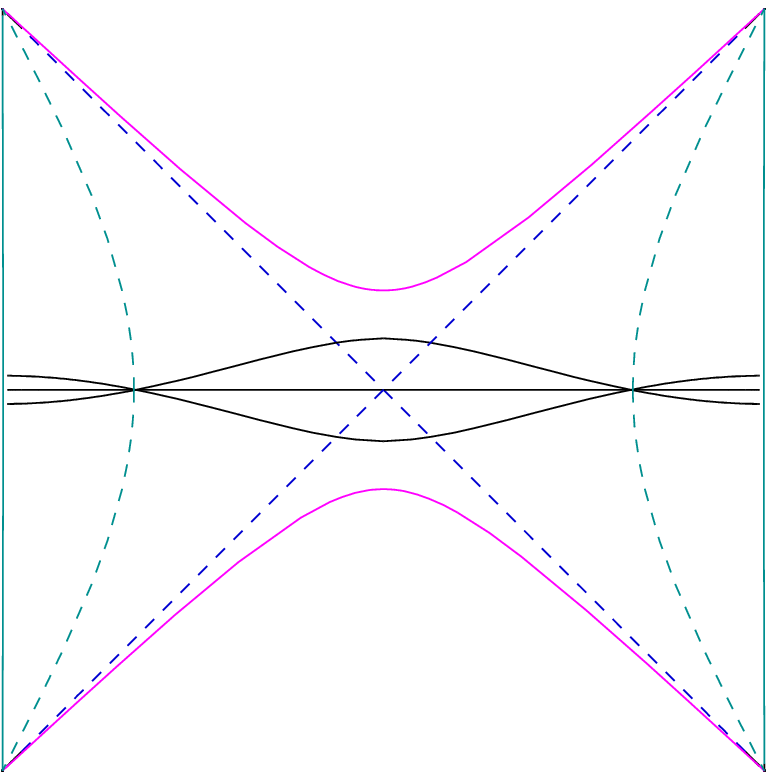}}

To track these branches, it is useful to resolve the branch point.
To do this we take $\rinf$ finite.  In the gauge theory this
corresponds to a finite UV cutoff.  Equations \Eexpans\ become
\eqn\Eexpanstwo{\eqalign{ t & \sim E^3 - a E \cr \len & \sim - E^4
- b E^2 ~.}} Here $a, b > 0$ are constants that go to zero as
$\rinf \rightarrow \infty$. At time $t=0$ there are now three real
solutions for $E$
and $\len$ (see \cubic). 
These correspond to the three geodesics shown in \threegeods.

We can most precisely define gauge theory correlators by
evaluating them in Euclidean time and then analytically continuing
to general time. To compute the  $t=0$ two sided correlator this
way we smoothly shift one point of a coincident point Euclidean
correlator by a half period, $i \beta/2$.   At $t=0$ we can do the
geodesic computation entirely in the Euclidean \sads\ geometry.
The correct geodesic is clearly the $E=0$ solution of \Eexpanstwo\
which is described in \length  .

Now let us go to small Lorentzian time by increasing the real part
of $t$ by a small amount.  The correct geodesic here is certainly
the small deformation of the $E=0$ solution with $E \sim t/a$.
This corresponds to following the central branch of the cubic
curve in \cubic.  But at a time $t \sim a^{3/2}$ this root of the
cubic annihilates with another root at the local maximum shown in
\cubic.  These solutions then become complex.  Complex values of
$E$ correspond to geodesics in the complexified \sads\ spacetime,
where both $r$ and $t$ are complex.  These can be described as the
solutions of the  equations of motion of the mechanics problem
with complex energy $E$.

Therefore the gauge theory answer for the correlator, defined by
continuation from Euclidean space, must become complex after a
certain time.  The complex branches of \longeqn\ do not contain
the $t_c$ singularity, and so this analysis explains its absence
in the gauge theory.  On the complex branches the second derivative of the real part of
$\len(t)$ is negative at $t=0$. In other words, the correlation
function computed by following this branch will decrease as we
move away from $t=0$, consistent with the general field theory
result.

\subsec{Finite mass black hole}

To understand the pattern of geodesics better, we now describe how to move the
bifurcation point to Euclidean time.  The advantage of studying
the Euclidean correlator is that, in the sum over paths
representation, the weight factor is always positive and so we can
reliably predict that the geodesics with the smallest $\len$
dominate. To  move the bifurcation we consider the finite mass
\sads\ geometry and compute $t(E)$ for its geodesics. The result
for small $E$ takes the form (Appendix D):

\eqn\Eexpansthree{\eqalign{ t &\sim E^3 - a \, E \cr \len &\sim -
E^4 - b \, E^2 ~.}} Here $a < 0$, unlike the finite $\rinf$
resolution described by \Eexpanstwo. The parameters $a$ and $b$ go
to zero as the black hole mass becomes infinite.   The parameter
$b$ is still positive.

 If $t= t_L + i \, t_E$ is Euclidean  ($t_L = 0$),
 then $E$ is pure imaginary. Let $E = - i \, {\tilde E}$,
with ${\tilde E}$ real.   Equation \Eexpansthree\ becomes

\eqn\Eexpansfour{\eqalign{ t_E &\sim {\tilde E}^3 - |a| \, {\tilde E}
\cr \len &\sim - {\tilde E}^4 - |b| \, {\tilde E}^2 ~.}}

\ifig\euclidgeo{Geodesics of finite mass Euclidean \sads\ with
finite $\rinf$.  The ${\tilde E}=0$ geodesic goes over the top and
has larger $\len$ than the two geodesics with nonzero ${\tilde
E}$.} {\epsfxsize=7cm \epsfysize=7cm \epsfbox{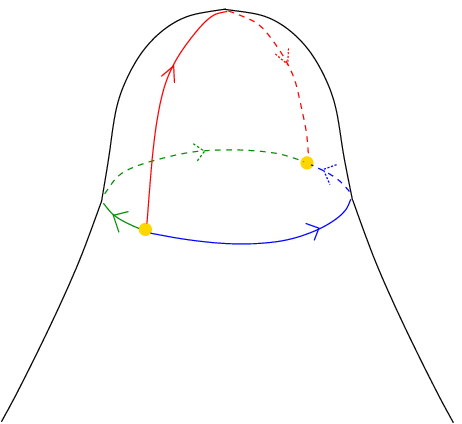}}

This cubic is also described by \cubic\ with the horizontal axis
now being ${\tilde E}$.   At large ${\tilde E}$, the extension of
\longeqn\ gives  $t_E \sim  \beta/2$ describing almost coincident
Euclidean points.  As ${\tilde E}$ is decreased, entering the
domain of validity of \Eexpansfour, a single solution exists,
describing a single geodesic.  But at a particular value of $t_E
\sim |a|^{3/2}$ corresponding to the local maximum of the cubic in
\cubic, a pair of geodesic solutions is created. From
\Eexpansfour\ and the sign of $b$ it is clear that these new
geodesics have larger $\len$ than the original one, so they are
exponentially subdominant.   This situation persists until $t_E
=0$, when the geodesic endpoints are antipodal.   Here the three
solutions of \Eexpansfour\ are ${\tilde E} = 0, {\tilde E} = \pm
\sqrt{|a|}$.  The two geodesics with ${\tilde E} = \pm \sqrt{|a|}$
have the same $\len$, which is less than that of the ${\tilde E} =
0$ geodesic.  This situation is illustrated schematically in
\euclidgeo\ where we have taken $\rinf$ finite for clarity. It is
clear from \euclidgeo\ that the ${\tilde E} = 0$ geodesic is a
local maximum of proper distance.

At $t_E = 0$ the correct prescription to compute the correlator is
to sum over the two dominant (${\tilde E} \neq 0$) geodesics.  We
then can continue into the Lorentz section by  varying $t_L$
away from zero.   To compute the correlator we follow both of
the local minimum geodesics. From \Eexpansthree\ we see that $E$ for this pair
of geodesics will now have to have nonvanishing real and imaginary
parts.  If we label the two geodesics $1$ and $2$, we have
$E_1=E_2^*$ and $\len_1 = \len_2^*$.  The real parts of the proper
distances are the same, so the magnitude of their contribution is
the same. So our final formula for the two sided correlator as a
function of Lorentzian time (or more generally, in a complexified
neighborhood of Lorentzian time) is given by \eqn\minkcorr{\langle
\phi \phi \rangle(t) = e^{-m\len_1} + e^{-m\len_2}.}

Because we have resolved the geodesic bifurcations we have been
able to give an unambiguous calculation of this correlator. Taking
the mass of the black hole to infinity we see that the correct
prescription using \longeqn\ is to sum over both complex branches
with equal weight.

 \subsec{A bifurcation model}

 As we have seen, at the branch
point described in \Eexpans, the geodesics coalesce.  This means
that stationary points of the path integral action become
arbitrarily close in path space.  This indicates that an
eigenvalue of the fluctuation operator around such a geodesic must
go to zero. There is a soft mode in the fluctuation spectrum
around the branch point geodesic. More explicitly, the soft mode
corresponds to infinitesimal motions of $E$ away from zero;
because $t(E) \sim E^3$ this means that the boundary points are
fixed under such motions.  In the space of paths, one can argue
that the coordinate $E$ is a good nonsingular coordinate for this
mode near zero.  For example, the central point of the path is
deflected by a distance proportional to $E$. Since the regularized
length behaves like $\len \sim E^4$ we see that the fluctuation
eigenvalue goes like $d^2 \len/ d E^2 \sim E^2 \sim t^{2/3}$. This
zero mode at $t=0$ will create divergences in the $1/m$ expansion
there.

The fluctuation dynamics near $t=0$ should be dominated by this
zero mode.   So we should be able to get a reliable picture of the
correlator in this region by truncating the path integral to this
one fluctuating mode.   Of course it will be non-gaussian, but
from our knowledge of $\len(E)$ we should be able to determine its
dynamics.   Representing the nonsingular coordinate for the zero
mode $E$ by $x = i E$, and setting $\tau=-it$ we can model the
path integral for $t \sim 0$ by the ordinary integral\foot{This
integral is a special case of Pearcey's Integral \BerryMII .}

\eqn\f{I(\tau) = \int_{-\infty}^{\infty} e^{-m V_{\tau}(x)} dx~,}
where the truncated action is given by \eqn\action{V_{\tau} =
\quarter x^4  - \tau  \, x.}

 To motivate this choice for $V_{\tau}$, compute its stationary points.

 \eqn\vsad{\eqalign{V_{\tau}'(x_s) &= x_s^3  - \tau =
 0 \cr x_s &= \tau^{\third} \cr V(x_s) &=-{3 \over
4} \tau^\ft~.}}

This is the analog of \Eexpans.\foot{ We will be able to study
this integral in sufficient detail that we will not need to add
the additional terms $\sim x^2$ to $V$ that would resolve the
cubic branch point.  This can be done though.}
 For $\tau$ real, ($t$ Euclidean), the integrand in \f\ is real and
 positive on the real $x$ integration contour.   In such a
 situation the large $m$ saddle point analysis is straightforward.
 We can then study deformations of $t$ away from Euclidean.

We now  study this model for complex $\tau$. We can still use
saddle point approximation for large $m$, but now we have to
determine carefully which saddles lie on the integration contour.
More precisely, for complex $t$ we must deform the integration
contour so that it becomes a steepest descent path for the real
part of $V_{\tau}$.  Then we can evaluate $I(\tau)$ by doing a
saddle point expansion around all the saddles included on the
path. A steepest descent path for the real part of $V_{\tau}$ is
simply a contour of constant $\Im (V_{\tau})$, so they are easy to
identify.

\ifig\stokesa{Integration contours for various values of $t$: left
to right, $\tau = 1, e^{i \pi /8}, e^{i \pi/5}$.  At $\tau = e^{i
\pi /8}$, the integral crosses a Stokes line and picks up another
saddle.} {\epsfxsize=13.5cm \epsfysize=3.5cm
\epsfbox{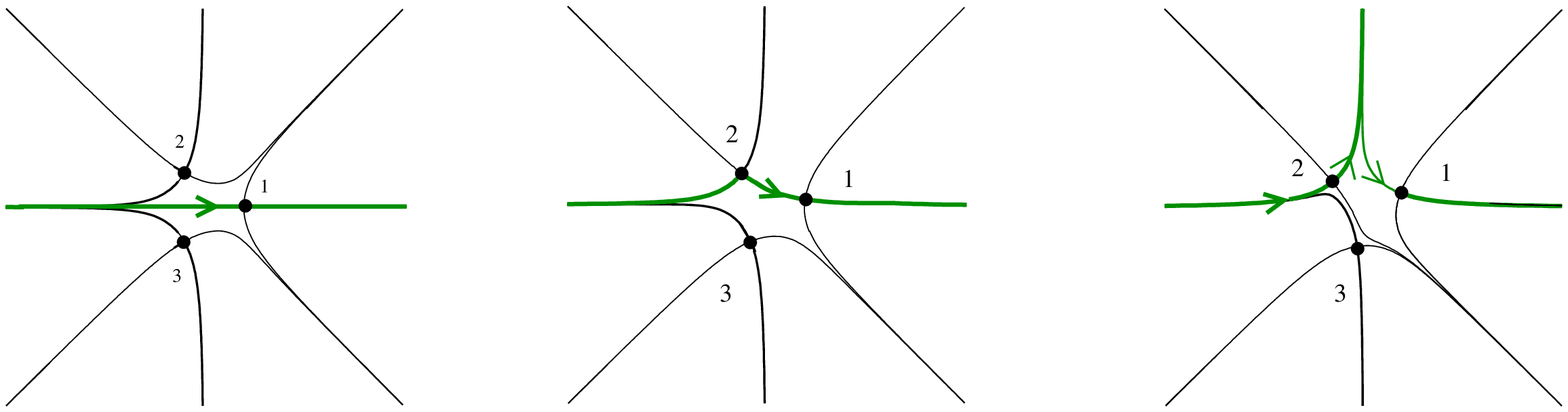}}

\ifig\stokesb{Positions of the Stokes and anti-Stokes (saddle
exchange) lines in the complex $\tau$ plane.  The dashed lines are
the Stokes lines, where the contour goes from crossing one saddle
to two, and therefore picks up a sub-dominant contribution. The
two saddles contribute equally on the wavy lines, so that the
integral there has a kink (in the large $m$ limit) as a function
of $\tau$.}
{\epsfxsize=8cm \epsfysize=5.8cm \epsfbox{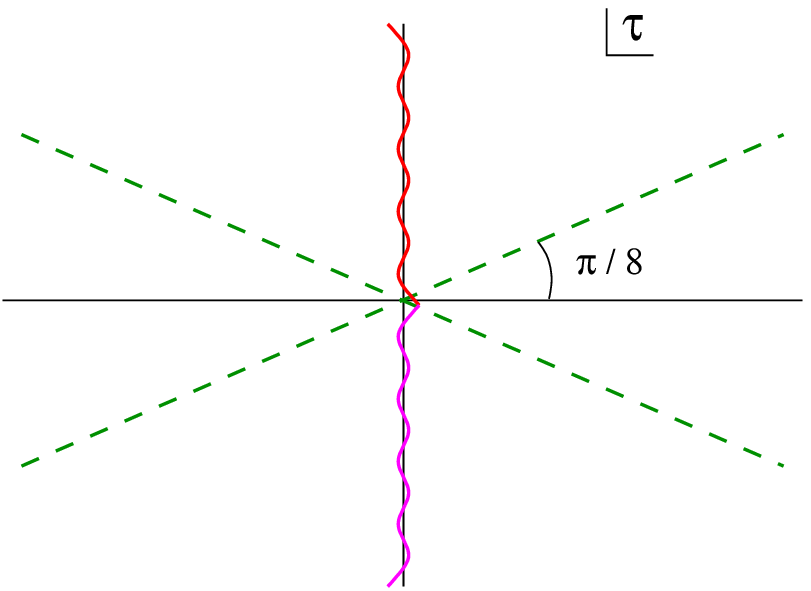}}

For convenience, we restrict ourselves to $\Im (\tau) \geq 0$, and
write \eqn\definet{\tau = r \, e ^ {3 i \theta}} with $\theta \in
[0, {2 \pi \over 3}], r \geq 0$.  The critical points are then
\eqn\criticalp{x_k = r^{\third} \, e^{i \theta + {2 \pi i k \over
3}}} where $k = 0, 1, 2$.  Consider starting with real positive
$\tau$, and then rotating $\tau$ counterclockwise into the complex
plane.  For $3 \theta < {\pi \over 8}$, the steepest descent
contour contains only the saddle point $x_0$.  For example, at
$\theta = 0$, the steepest descent contour is simply the defining
contour of the integral:  the real axis. At $3 \theta = {\pi \over
8}$, however, the imaginary parts of $V_{\tau}$ become equal at
$x_0$ and $x_1$, and the contour picks up the saddle point $x_1$
(\stokesa).  The line $3 \theta = \pi/8$---where the integral
picks up a subdominant exponential in the saddle
approximation---is known as a Stokes line \BerryMI.  This
situation persists until $3 \theta = {7 \over 8} \pi$, at which
point the contour loses the point $x_0$, and only $x_1$ remains
(\stokesb).

By computing the real part of $V_{\tau}$, we see that the saddle
at $x_0$ dominates over $x_1$ for $3 \theta < {\pi \over 2}$, and
vice versa for $3 \theta > {\pi \over 2}$.  The locus of this
exchange of dominance is referred to as an anti-Stokes line. Right
on the line both saddles contribute equally.   This is the
situation for the Lorentzian correlator discussed in the previous
subsection.

An interesting phenomenon occurs here.  For ${\pi \over 4} < 3
\theta < {3 \pi \over 4}$, the overall dominant saddle is $x_3$,
which is not on the integration contour, and does not contribute
to the integral in this approximation. This is similar to the
phenomenon we encountered in the \sads\ correlator, where the
dominant branch (the $t_c$ geodesic) does not contribute.  That we
have a situation in which a dominant saddle does not contribute to
the contour integral may at first seem surprising.   In fact, it
is easy to see that no matter how one tries to deform  the contour
so as to force it to go over this dominant saddle in  a meaningful
direction, one will pick up a larger extremum elsewhere on the
contour.  This extremum is of course only an extremum for
$V_{\tau}$ restricted to the contour. It is not a critical point
of $V_{\tau}$ as a function of complex $\tau$.  However we do not
have an analogous understanding of why the $t_c$ geodesic does not
contribute to the Lorentzian functional integral.  There must be
other large contributions that cancel it.

For $m \rightarrow \infty$ subdominant saddles make no
contribution to this analytic function.  The true $ m \rightarrow
\infty$ correlation function \foot{More precisely, we should
consider the function $L(\tau)=-{1 \over m} \ln I(\tau)$ which has
a smooth large $m$ limit.} is a set of analytic regions, joined
together in a non-smooth way at anti-Stokes lines when saddles
exchange dominance. This is somewhat analogous to a first order
transition in statistical mechanics. Computing the correlation
function in one region and analytically continuing ignores these
boundary lines, and is somewhat like continuing into a metastable
phase. This analogy is imperfect, though.  The bifurcation model
illustrates that the $t_c$ saddle does not give a contribution
like $e^{-m\len}$ to the path integral as would a metastable
phase.

Now consider $I(\tau)$ for large but finite $m$.   It is clear
from the defining integral \f\ that $I$ is an entire function of
$\tau$.  The $m = \infty$ branch point at $\tau=0$ is smoothed
out. The integral has an everywhere convergent power series and in
a small disk $|\tau| \ll 1/m^{3 \over 4}$, $I$ is roughly
constant. When  $|\tau| \sim 1$ the saddle point approximation is
very accurate for large $m$.   Along the anti-Stokes (saddle
exchange) line, the vertical axis in \stokesb , there is a rapid
but smooth transition from one saddle behavior to the other.   The
width of this transition is order $1/m$.  So there are two half
plane regions where the large $m$ saddle point behavior is very
accurate, a thin joining region along the anti-Stokes line and a
small roughly constant disk near the origin.  These features
smooth out the branch cut and so prevent us from following
subdominant saddles by analytic continuation.

A concrete method for performing analytic continuation is by
computing power series coefficients around a point and then
continuing the series variable.   If we expand, for example,
around a point away from the anti-Stokes (saddle exchange) line at
large but finite $m$ the contamination of the subdominant saddle
point contribution that prevents continuation past the point of
saddle exchange is of order $e^{-m}$.  Near the exchange line we
can model the correction more accurately as $e^{-m(1-\tau)}$ where
the exchange line is at $\tau =1$.   The power series expansion of
this correction goes like $e^{-m}\sum (m \tau)^k/k!$ .  For fixed
$k$ the corrections are order $e^{-m}$ but for $k \sim m$ the
corrections are order one.    So the practical question about how
large an $m$ is required to be effectively infinite is equivalent
to how many terms of a power series  are required to accurately
represent the infinite $m$ function.   The functions we are
interested in have smooth large $m$ limits so the number of terms
required is $m$ independent.   This means that accurate analytic
continuation can be done with results at large but finite $m$.
Examples of this procedure for studying the $t_c$ singularity are
discussed in Appendix F.

There is an asymptotic series of $1/m$ corrections to $I(\tau)$
that can be computed by Feynman diagram techniques.  Because of
the soft mode, each of these terms is divergent at $\tau = 0$ and
has the structure $1/(m \tau^{4 \over 3})^k$.  So each term lives
on the same three sheeted Riemann surface as the leading answer.
If we compute a given term around a dominant saddle and then
analytically continue to a region where that saddle is no longer
dominant we find the same answer we would get by perturbing around
the subdominant saddle using a different integration contour.  So
the fluctuations we compute around the dominant saddle give us
information about fluctuations around the subdominant one.   We
can clearly do the same thing with corrections in powers of $a$.
We will use generalizations of this idea to argue that we can
study stringy and quantum effects around the bouncing geodesic.

\subsec{The $t_c$ singularity}

We now have a reliable calculation of the boundary correlator at
large $m$ in the classical supergravity approximation as a
function of complex boundary time $t$.   The result is contained
in Equation \longeqn, with additional information about which
geodesics dominate.   For Lorentzian time $t= t_L$  we sum over
both complex solutions of \longeqn\ with equal weight.   The
complex branches of \longeqn\ are free of additional singularities
for all finite $t_L$.  But the functions $\len_i(t)$ are analytic
in $t$. Given their values along any small segment of $t$ it is
possible to analytically continue them over the entire Riemann
surface described by \longeqn. The $t_c$ singularity can be found
by such an analytic continuation from the dominant branch selected
by the boundary CFT. In the discussion above of the bifurcation
model we started from Euclidean $t$. Here let us give the argument
starting from the Lorentzian  $t$.  Consider small $t$ and hence
use the expansion \Eexpans, which gives us the relation $E \sim
t^{\third}$. As we have argued previously, for small Lorentzian
$t$ the boundary CFT selects the branch $E= e^{{2 \pi i \over 3}}
|t^{\third}|$. (We suppress the other complex branch which behaves
similarly.) If we denote $t=e^{i\alpha} t_L$ then we reach the
branch with real $t$ and real $E$ (and hence real $\len$) by
taking $\alpha =\pi$.  On this branch there is a singularity at
$|t| = t_c$, due to the logarithmically diverging $\len$ of the
analytically continued geodesic, which ``bounces" off the
singularity. This $t_c$ singularity clearly reflects aspects of
the black hole singularity. So we see that information about the
black hole singularity  is encoded in the boundary correlators,
albeit in a subtle way involving analytic continuation.   Despite
its subtlety, analytic continuation can be computationally
effective. In Appendix F we show how a modest number of terms in a
power series expansion of $\len$ on a dominant branch is enough to
extract precision information about the $t_c$ singularity.

\subsec{Quasinormal modes}

\ifig\timecontour{Contours of constant
imaginary part for the function $t(E)$.  There are three $\Im(t) = 0$ contours,
which cross at the origin.  The $t_c$ branch is the real axis.  The other two give us
the non-$t_c$ branches of the correlator.  Pane b) is a magnified
view of the singularity at $1+i$, showing only the $\Im(t)=0$ contour.}
{\epsfxsize=12cm \epsfysize=5.8cm \epsfbox{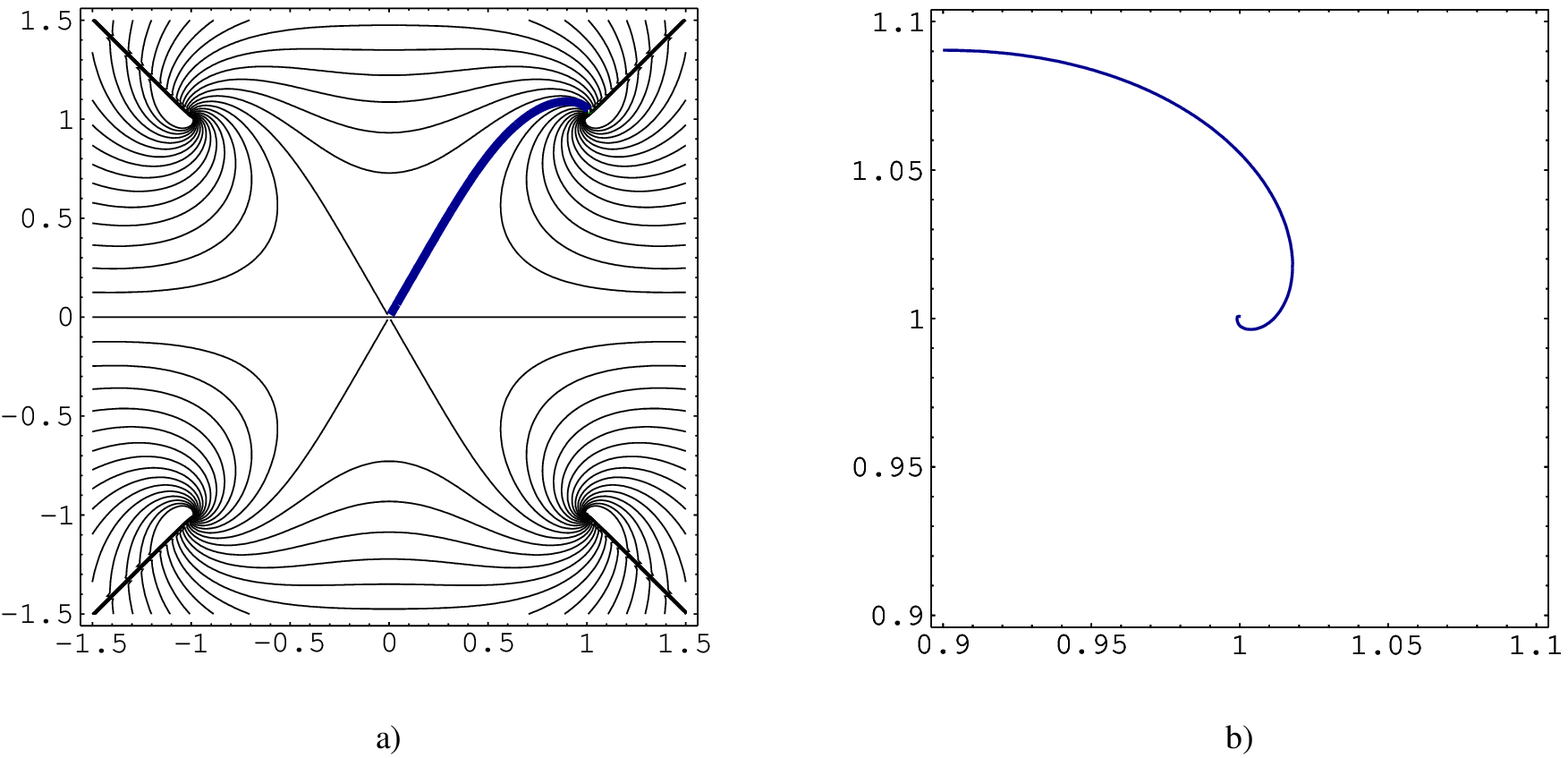}}

We can use the geodesic expression for the correlator derived
above to compute the quasinormal modes of the \sads\ black hole,
in the large $m$ limit.  These agree with the known results
\NunezEQ, providing a check of our calculation. \foot{We thank Chris Herzog 
for pointing out a mistake in a previous version of this section.}

To compute quasinormal modes we need the large $t$ asymptotics of
the correlator.  We first note that along either complex branch
$E(t)$ must go to one of the fourth roots of $-4$ as $t$ goes to
infinity. Specifically, let $\a_1$ and $\a_2$ be $-1-i$ and $1-i$
respectively; we can rewrite formula $\longeqn$ as \eqn\longeqna{t
= {1 \over 4} \ln \[ (E - \a_1) (E - \bar{\a_1}) \over (E - \a_2)
(E - \bar{\a_2})\] - {1 \over 4} i \ln
\[ - (E - \a_1) (E - \a_2) \over (E - \bar{\a_1}) (E - \bar{\a_2})
\].}  We can expand around these
various roots; for example, expanding $E$ around $\a_2$ gives $t =
c - \( \a_2 / 4 \) \ln (E - \a_2)$, where $c$ is a complex
constant.  Writing $E - \a_2 = r e^{i \theta}$ and requiring that
$t$ be real gives \eqn\treal{ \theta = \ln r + c',} where $c'$ is
a real constant.  This demonstrates that the complex branches
(where $t$ goes to real infinity) are spirals in the complex $E$
plane, converging on one of the roots (see \timecontour). Coupled
with the expansion $\len \sim - {\half} \ln (E - \alpha)$ around
any of the four roots, we have that as $t$ goes to infinity, $t
\sim \half (1 + i) \len$ along one of the complex branches and $t
\sim \half (1-i) \len$ along the other. This gives the leading
large $m$ behavior of $e^{-m \len_1} + e^{-m \len_2} = e^{- m(1+i)
t} + e^{- m(1-i) t}$.  This reproduces the first large $m$ quasinormal
modes of this black hole found in \NunezEQ.  Note that conventions differ 
between \NunezEQ\ and this paper.

To go beyond the first quasinormal mode, we argue as follows.
Suppose we are on the complex branch that goes to $\a_2$ as $t$
goes to infinity.  From $\lengthEr$ we then see that $e^{-\len} =
{(E - \a_2)}^{\half} \tilde{f} (E)$, where $\tilde{f}$ is
analytic. We can invert and obtain \eqn\qa{E = \a_2 + e^{-2 \len}
f(e^{-2\len})} with $f(0) = 1$ and $f$ analytic.  Substituting, we
obtain  $t = {\half} (1+i) \len + \tilde{g}(e^{-2 \len})$, with
$\tilde{g}$ analytic.  Inverting once more we obtain \eqn\qb{\len
= (1-i) t + g(e^{-2 (1-i) t}).} Now, the correlator is
\eqn\qba{\langle\phi\phi\rangle(t) = e^{-m (\len(t) + (1/m) f_1
(t) + \cdots)}} where $f_1$ is the first $1/m$ correction.  By
extensivity and locality, we know that $f_1(t) = b_0 t + h(t)$
where $b_0$ is a constant of order one determined by the
fluctuation determinant around the geodesic (which we have not
calculated) and $h(t)$ is exponentially small in $t$ because the
world line path integral is massive in \sads. Expanding gives a
series

 \eqn\qc{\langle\phi\phi\rangle(t) \sim
 \sum_{n=0}^{\infty} c_n e^{-(m+b_0+2n)(1-i)t } ~~ + {\rm c.c.}}
This yields values for the quasinormal modes \eqn\qom{\omega_n= (m
+ b_0+ 2n)(1\pm i)~~~n=0,1,2 \ldots ~.} These values agree with the
large $m$ quasinormal modes calculated in \NunezEQ\ if $b_0=
-1$.  The results in \NunezEQ\ were obtained 
by a direct study of the radial equation for the \sads\ geometry, the Heun
differential equation. In some way the geodesic computations
summarized in \longeqn\ give a WKB solution to the Heun equation.

\newsec{Beyond the geodesic approximation}

In the previous section we studied the massive scalar correlator
in the limit $ m \rightarrow \infty$ where the geodesic
approximation applies. We worked in the classical supergravity
regime, $\alpha ' = l_s^2 = 0$ and $g_s=0$. In this section we
will discuss the more general situation.

\subsec{Finite mass}

The scalar propagator has a path integral representation
\eqn\pathpropscale{ \langle \phi(\xi_1)\phi(\xi_2) \rangle =
\int_0^{\infty} dT ~\int
 Dx(s) \exp\left(-i m/2 \int_0^T ds ~
 \left( {\dot x}^2 +1 \right) \right).
   }
Here $\xi_i$ are points on the boundary and the integral is over
paths in the bulk that connect these two points.  For the radially
separated points discussed in the previous sections $\langle
\phi(\xi_1)\phi(\xi_2) \rangle = \langle \phi \phi \rangle (t)$.
The mass $m$ acts like $1/\hbar$, a saddle point parameter. Taking
$m$  finite means all paths are explored in the functional
integral, not just the dominant saddles.   So one of the key steps
of the previous section--following a saddle point into a region
where it is no longer dominant--will be problematic when $m$ is
finite. More explicitly, as discussed in section 6.5,  the
function $L(m, t) = -{1 \over m} \log \langle \phi\phi \rangle(t)
$ defines a piecewise analytic function in the limit $m
\rightarrow \infty$ that can be continued past the anti-Stokes
line where saddle point dominance is exchanged, but at finite $m$
there is no clear way to follow the subdominant saddle past this
line. Operationally, the ``contamination" due to subdominant
saddles must be controllably small to perform such analytic
continuation.

The coefficients of the $1/m$ expansion  can be continued
unambiguously and do contain significant physical information.
 From \pathpropscale\ we see that this expansion is a small
fluctuation or heat kernel expansion.  There are two places where
divergences are expected in the coefficients of $(1/m)^k$.  The
first is $t=0$, where multiple geodesics coincide. As discussed in
section 3.3 we expect a soft mode in the fluctuations around the
geodesic here.  This will cause a divergence in the $1/m$
expansion. The terms in the $1/m$ expansion live on the same three
sheeted Riemann surface as $\len$.  As discussed above, continuing
these coefficients gives the small fluctuation expansion around
the analytically continued saddle.

The second place we expect divergences is on the bouncing geodesic
for $t \sim t_c$.  Here the divergence is not due to a soft mode
but to large nonlinearities present because the curvature is large
near the singularity.  We can expand the heat kernel
\pathpropscale\ in a short time expansion \eqn\shorttime{ \langle
\phi(\eta_1)\phi(\eta_2) \rangle \sim \sum_{k=0}^{\infty} c_k
{\mit R}^{(k)} \left({T \over m} \right)^k, } where $\eta_i$ are
bulk points near the singularity lying on the bouncing geodesic
and ${\mit R}^{(k)}$ is shorthand for a curvature invariant that
has scaling dimension $({\rm length})^{-2k}$. Because the only
length scale near the singularity is ${\mit l}$, the shortest
proper time between the geodesic and the singularity, we have
${\mit R^{(k)}} \sim {\mit l}^{-2k}$. The region of proper
distance $T$ where the geodesic is near the singularity is also
$\sim {\mit l}$. Putting in these estimates, \shorttime\ becomes
\eqn\shorttimel{ \langle \phi(\eta_1)\phi(\eta_2) \rangle \sim
\sum_{k=0}^{\infty} c_k \left( {1 \over m \, {\mit l} } \right)^k
~.} This is reasonable since $m \, {\mit l}$ is the only
dimensionless combination available near the singularity. From
\tdiff\ we see that $r_{min} \sim 1 / E$ and $E \sim 1/(t-t_c)$.
It follows immediately from the metric that ${\mit l} \sim
r_{min}^2 \sim (t-t_c)^2$. So we expect the $1/m$ expansion to
have the structure \eqn\shorttimee{ \langle \phi\phi \rangle(t)
\sim \sum_{k=0}^{\infty} c_k \left( {1 \over m \, (t-t_c)^2
}\right)^k ~.}  We should be able to compute these coefficient
functions on the primary sheet and the analytically continue them
to the secondary sheet to study their singular behavior. This
gives an example of recovering nontrivial information about the
neighborhood of the singularity from outside the horizon
correlators. The information here is just the diverging curvature
near the singularity, and it is clear from the above that the main
contribution to the singular behavior comes from distances $\sim
{\mit l}$ which can be made arbitrarily short (in supergravity
approximation) by taking $t \rightarrow t_c$. It is possible that
the strengthening of the singularity with increasing $k$ will have
a recognizable signature in the correlator on the first sheet.

\subsec{Finite $\alpha'$}

We now go beyond the supergravity approximation by letting the
string length $l_s$, ($ ~\alpha' \sim l_s^2$ ) be finite.  In the
boundary SYM  theory, this corresponds to taking the 't Hooft
coupling $\lambda = g_{\rm YM}^2 N = (R/l_s)^4$ finite. We still
work in the classical limit $g_s \rightarrow 0, ~ N \rightarrow
\infty$.

When $l_s$ is finite, the behavior of correlators in the large $m$
limit depends on which large $m$ excitation we are studying.  If
we take a generic perturbative string state with many oscillator
modes excited then the size of the excitation will diverge as $m
\rightarrow \infty$ and the correlator will cease to serve as a
local probe.  Instead we consider D-branes, which, at $g_s =0$ are
essentially pointlike, infinitely massive objects \refs{\ShenkerXQ
, \KabatCU , \DouglasYP} . In particular in the $AdS_5 \times S^5$
IIB string theory we study D3-branes wrapped around an $S^3$
submanifold of $S^5$. Such a state is pointlike in the $AdS_5$
space and realizes the BPS state carrying SO(6) charge, for large
charge.   These are the giant gravitons \McGreevyCW.

The main effect of small but finite $l_s$ on such states is
encapsulated in corrections to the supergravity Lagrangian.  These
include, for instance,  corrections of the form $(\alpha')^k
R^{(k)}$. The small parameter controlling the size of such terms
is $l_s/{\mit l}$. For phenomena near the horizon ${\mit l} \sim
R$ and this small parameter is of order $l_s/R \sim
1/\lambda^{1/4}$. Such terms have a number of effects on the
dynamics, including small shifts of the metric and dilaton fields
\GubserNZ.

We can estimate the effect of such shifts on the geodesic dynamics
by following the discussion in the previous section about the
finite mass black hole.  Just as in that situation, the most
general small, smooth shift in the geodesic equations can be
encapsulated by deforming \longeqn\ and \lengthEr\ into
 \eqn\edeform{\eqalign{t &= E^3 - a E    \cr
  \len &= -E^4 -b E^2  }}
Here $a, b$ are small parameters whose signs are undetermined. As
in the discussion following \Eexpanstwo\ and \Eexpans\ this
structure guarantees that the bifurcation, needed for analytic continuation,
 does not disappear at small but finite $l_s$. At
most the cube root singularity at $t=0$ splits into two square
root singularities.

This persistence of the singularity at small $t$ creates a puzzle.
Explicit field theory calculation of the two point correlator at
small $\lambda$ (i.e., large $l_s$) shows no sign of such a
singularity. So it seems there must be an $m = \infty$ phase
transition at finite $\lambda$.   Perhaps we have missed some
effect at large $\lambda$ that removes the singularity.  On the
other hand it does not look difficult to construct metric and
dilaton modifications that remove the singularity for a finite
size deformation.

The nature of $l_s$ corrections for $t$ near $t_c$ gives
information about  stringy behavior near the black hole
singularity.  As discussed above, we expect corrections to be
controlled by the dimensionless ratio $l_s/{\mit l} =
\lambda^{1/4}/(t-t_c)^2$.  So there should be large corrections
for $t \sim t_c$.

Of course, this is a string tree level effect and can in principle
be computed by string world sheet techniques.  The D-brane
trajectory is encoded in a boundary state.  The analytic
continuation necessary to study the real geodesic requires
boundary states with complex D-brane positions.  Such states have
been studied recently in other contexts \LambertZR\ \GaiottoRM.
Perhaps one can circumvent the analytic continuation by directly
formulating the conformal field theory in Lorentzian space, and then
selecting the boundary state corresponding to a D-brane following
the real geodesic.

\subsec{Finite $g_s$}

We now relax the last remaining constraint and consider finite
$g_s$.  A problem immediately arises because the masses of
D-branes are $\sim 1/g_s$ and so the mass $m$ of the probe
particle cannot be taken infinitely large. More precisely if the
$S^3$ radius on which the $D3$-brane is wrapped is $\sim R $, then
its mass $m$ is $ m R \sim (m_s R)^4/g_s \sim \lambda/g_s \sim N$.
If the $S^3$ radius is string scale then $m R \sim m_s R/g_s \sim
N^{1/4}$.  If the SO(6) charge is $Q$ then $m R = Q$.\foot{This is
only approximately true because supersymmetry is broken in the
black hole background, and the BPS conditions no longer strictly apply.  But the
semiclassical analysis in \McGreevyCW\ demonstrates that this
relation continues to be approximately true for large $Q$.} The
stringy exclusion principle \MaldacenaBW\ and the giant graviton
analysis reviewed above show that stability requires $Q \le N$. As
discussed above, at finite $m$ the correlation function
contribution from the dominant saddle is ``contaminated" by
$e^{-m}$ corrections from subdominant saddles that will eventually
become dominant on crossing an anti-Stokes line.  We can certainly
study correlators as a power series in $g_s$ since $g_s^k \gg
e^{-m} \sim e^{-\lambda/g_s}$ as $g_s \rightarrow 0 $. Because the
emission and absorption of virtual perturbative quanta cannot
significantly alter the D-brane trajectory, we should be able to
analytically continue the geodesic through the anti-Stokes line.
   This will allow us, in principle, to study the
nature of string perturbation theory in the vicinity of the
singularity by continuing each order across the anti-Stokes lines
to $t \rightarrow t_c$. Again, these quantities should in
principle be computable by studying higher genus string diagrams
with the appropriate boundary state.   Of course there will be
$1/m$ corrections to these  order $g_s^k$ amplitudes which can be
numerically larger than higher order $g_s$ corrections. But these
are fluctuations about the main saddle, and will each analytically
continue to the desired real geodesic.    They can be
distinguished from $g_s$ effects by their different parametric
dependence on $\lambda$, or equivalently $Q$.

In the supergravity approximation at least, we might expect
divergences in loop amplitudes when interaction vertices approach
 the singularity.  This issue was discussed for $d=3$ in
\KrausIV. On general grounds, applicable here as well, such
divergences should not occur.   The correlators are manifestly
finite in Euclidean SYM.  They are analytic and so can be singular
at most on a set of complex codimension one.  So a generic
divergence is not possible. From the bulk point of view the
potential divergences in $d=3$ are not present because the
analytic continuation from Euclidean space induces a shift $r
\rightarrow r + i \epsilon$ which regulates the singularity.
Interaction points are integrated across the singularity into
another region to implement this.  Remaining imaginary parts are
cancelled between past and future singularities.  Although $d>3$
is more complicated we think it reasonable to expect a  similar
mechanism here.

But the elimination of divergences need not eliminate large finite
contributions.   Since the geodesic near $t_c$ is very close to
one singularity it is natural to expect large finite contributions
in quantum corrections.  We think this is what is being computed
by the boundary theory near $t_c$.

 We would like, at least in principle, to be able to obtain
nonperturbative information about the singularity from the
nonperturbatively well defined boundary CFT.

Operationally, the size of nonperturbative phenomena whose
continuation we want to study must be parametrically larger at the
point we calculate than the $e^{-m}$ contamination from the
subdominant saddles that would become dominant at an anti-Stokes
line and prevent analytic continuation of the ``metastable" phase.
For example, suppose we wish to study D-brane effects, for
instance the pair production of small D-branes.  Such effects are
typically of size $e^{-1/g_s}$.  Using the largest stable giant
graviton as a probe, $e^{-m} \sim e^{-\lambda/g_s}$. Since
$e^{-1/g_s}$ is parametrically larger than $ e^{-\lambda/g_s}$, it
should be possible to reliably compute such processes and then
analytically continue them to the real sheet.  Processes that may
be impossible to compute include NS brane production, expected to
be of order $e^{-1/g_s^2}$.  It may be possible, though, to
isolate these effects by studying the boundary theory as an
analytic function of $g_s$ and $\lambda$.

Processes where the nonperturbative effects are dominant should be
much easier to study.  As a first example, consider  D-instanton
effects that are the leading contribution to certain anomalous
processes involving the bulk axion, related to the boundary
$\theta$ parameter.   The effect of $\theta$ on giant graviton
correlators should behave like $e^{-1/g_s}(f_0(t) + O(e^{-m}))$.
Here we have normalized to the $\theta = 0$ answer. The function
$f_0(t)$ can in principle be computed  on the primary sheet with
parametrically small error. Then it can be continued to the
secondary, real sheet. If D-instanton effects are enhanced near
the singularity then there should be a signal in $f_0(t)$ near
$t_c$.

As a second example,
 consider a charge $Q$ wrapped D-brane ``particle" of mass $m \,
R = Q$. Take $Q \sim N$. Now imagine a process where this particle
fragments into two other such particles of charges $Q_1$ and
$Q_2$, with $Q_1+Q_2=Q$ and $Q_i \sim Q/2$. The black hole is not
a supersymmetric background so there is no BPS condition that
prevents this. But this process is certainly nonperturbative. By
considering the amount of D-brane that must be created and
destroyed we have provisionally estimated that the rate for this
process is $\sim \exp(-Q)$.\foot{In this entire discussion we have
not considered the D-brane wrapped on a sphere in ${\rm AdS_5}$
discussed \refs{\GrisaruZN,\HashimotoZP}. Such a state gets light
near the black hole singularity and so it should have a strong
effect on the analytically continued dynamics. It may be that the
nonperturbative tunneling to this configuration discussed in
\refs{\GrisaruZN,\HashimotoZP} is the dominant mechanism for
fragmentation near the black hole singularity.  This deserves more
study.} Both initial and final particles follow geodesics of the
kind we have been discussing, assuming that the final boundary
operators at placed at the same point. These geodesics should
continue smoothly onto the secondary sheet.   Schematically, we
expect the amplitude for this process to  look like
$e^{-Q}(f_0(t)+ O (e^{-Q}))$ where we have normalized to the two
particle charge $Q$ correlator. Again, $f_0(t)$ is in principle
calculable on the primary sheet with parametrically small error.
It can then be analytically continued to $t_c$.

The kind of quantum information that can be obtained results from
taking $g_s \rightarrow 0$ in a certain way.   If one focuses on
the behavior of amplitudes as $t \rightarrow t_c$ then perhaps the
information that can be extracted about the singularity is some
kind of double scaling limit that compensates for a shrinking
$g_s$ by examining processes that grow large for $t \rightarrow
t_c$.

\newsec{Discussion}

The analysis of the preceding sections shows that a significant
amount of information from behind the horizon, and in particular
from the region near the singularity, is encoded into boundary CFT
correlators.   An initial question we must consider is how this
information can emerge from the hot, thermal, apparently
featureless horizon.  Our understanding of this issue is limited,
but it is clear that analyticity plays a central role. It is
analyticity that lets us reliably follow a geodesic into a region
where its contribution is exponentially subdominant. One might
think that our results are merely a consequence of the analytic
nature of the classical \sads\ geometry, which allows one to
compute the complete metric from a small sample of it far outside
the horizon. But clearly more than this is involved. We have
argued in the previous section that a large amount of stringy and
quantum information should be accessible by analytic continuation
as well. It seems that it is the analyticity of the full quantum
theory, or at least a limiting part of it, that is 
the central issue.

Analyticity is not a central notion in classical general
relativity.  For instance, in the collapse of a matter shell to
form a black hole the metric is not analytic, and only one
asymptotic region exists.  But outside the shell the metric is
just the standard black hole metric.  Correlators will in general
not be analytic in such a spacetime.  It seems that our methods
may not apply to such cases.   Even with analytic initial data, the
same side correlators in a collapse scenario seem to differ from
those of the eternal black hole by exponentially small terms, even
though the geometry behind the horizon differs markedly.

However, analyticity does play a central role in quantum field theory.
Multipoint correlation functions are analytic functions of the
point locations.   So quantities computed in AdS/CFT will
generically be analytic.   We can envision setting up a collapse
scenario in AdS/CFT by applying a large number of boundary
operators at a fixed time far in the past.  We then let this
matter evolve.   From the boundary point of view, we  excite many
gluons  and then let them thermalize.   We have created one of the
microstates counted in the black hole entropy. We then can insert
two very massive particle boundary operators as probes. Their
correlator will be analytic in their time separation, and should
go over to the thermal ensemble in the limit of large energy,
which corresponds to large black hole mass. But even away from
this limit the probe correlator is precisely analytic. In
principle, we can test for periodicity in imaginary time, remnants
of the $t=0$ branch point and of the $t_c$ singularity.    We can
look for exponentially small one sided contributions that become
significant when continued to opposite sides.   It may well be
difficult to disentangle these small terms from the $e^{-m}$
contamination that obstructs analytic continuation.   The
different parametric dependence and ability to study different
microstates might make it possible to distinguish the two. This
question may be easier to study in $d=3$ where the real behind the horizon
geodesic dominates for all $t$.   If the effective geometry behind
the horizon is modified then it should make a large effect on this
two sided correlator.

These questions should be related to the issue of recovering
information that has  fallen into the black hole.  The lack of
information loss in eternal black holes is signalled by
Poincar{\'e} recurrences \MaldacenaKR, which occur on time scales
$t_r$ exponentially longer than $t_c$, $t_r \sim \exp(N^2) t_c$.
These recurrences may correspond to large fluctuations in the
classical geometry \MaldacenaKR. These occur at times $\sim N^2
t_c$ .  It would be very interesting to connect these ideas to
those discussed in this paper, although the enormous difference in
time scales makes this challenging.

 As mentioned in Section 3,  tracking the bouncing geodesic
through the anti-Stokes line is a bit like  following a metastable
phase in a statistical mechanical system, although this analogy is
imprecise.  A related phenomenon has already been discussed in the
AdS/CFT correspondence.  Gross and Ooguri \GrossGK\ considered the
expectation value of two facing Wilson loops of radius $R$ and
separation $L$.  In the limit of infinite string tension $\lambda
\rightarrow \infty$ this quantity is determined by the minimal
area string world sheet spanning the loops.  As $L$ is increased
this surface jumps in a discontinuous, ``first order" transition.
At large $L$ the disconnected surface dominates. But one can in
principle continue through this ``anti-Stokes line" and study the
metastable connected surface until it pinches off into a kind of
singularity. It would be interesting to consider various kinds of
fluctuation corrections in this situation and see which ones can
be continued through the anti-Stokes line to the pinch off point.

\ifig\bat{Symmetric spacelike geodesics of \sads.  The numbers
indicate how many such geodesics pass through each point in that
region.}{\epsfxsize=7cm \epsfysize=7cm \epsfbox{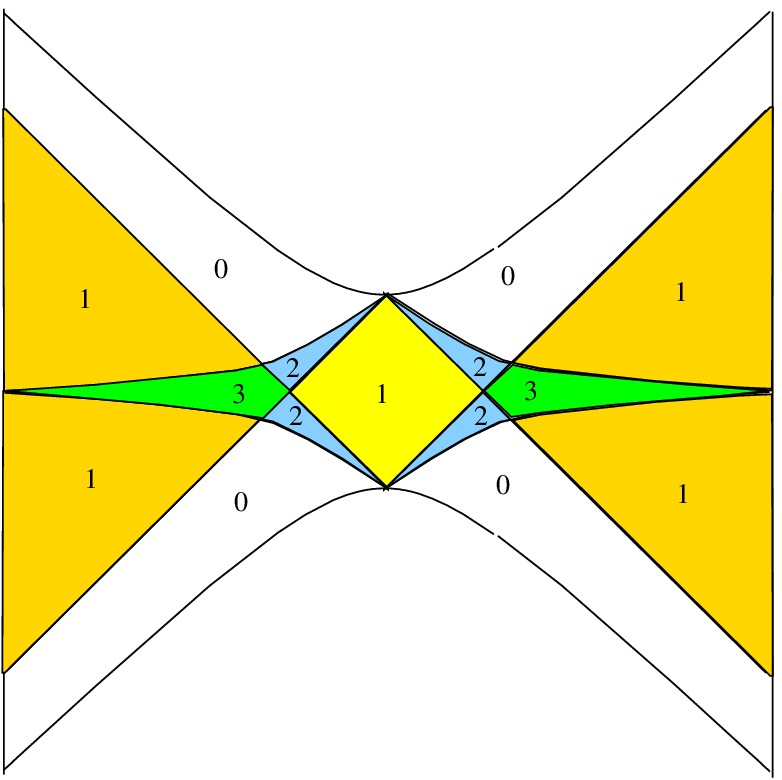}}

We have argued that the real geodesic (which   becomes almost null
at $t=t_c$), contributes to boundary correlators in a subtle
subleading way that can be exposed by analytic continuation to a
secondary sheet.   We might ask whether it ever provides the
dominant contribution to a physical quantity.  We are not sure of
the answer to this question. But, if we temporarily put aside the
well founded concerns about measuring local bulk correlation
functions in quantum gravity, and just compute them in classical
supergravity approximation, we can find some answers. In \bat\ we
have displayed the regions for which a symmetric bulk two point
function is dominated by various
geodesics. Inside 
regions 3, 2, and the central 1,
the most likely scenario is that the
geodesic that dominates the correlation function lies on the real
branch (real $E$) and so will evolve into the nearly null geodesic.  For the
dominant geodesic to get within a proper time ${\mit l} $ of the
singularity, the correlator points have to be separated by a proper
distance $\sim {\mit l}$.  For these quantities the large stringy
and quantum effects discussed in the previous section should be
important factors in computing the leading result, not a subtle
subleading contribution.  But of course the role of such bulk
quantities is a deeply confusing one.

 We now  ask what  this new information can tell us about
the singularity.  Unfortunately, little is known about boundary CFT
correlators at strong coupling.  In fact, our results in the bulk
make many new predictions
about them, as has often been the case in AdS/CFT.  Still, it is a
useful exercise to ask what could be learned if complete
correlator information were available for the CFT.

In previous sections we concentrated on the two point correlator.
But there is no reason not to study more general correlators (an
example of this was discussed in Section 4).  In particular, consider
the correlator with $K $ operators on the asymptotic boundary of
region I and $K'$ on the boundary of region III, with a variety of
$SO(6)$ charge assignments. Each operator corresponds to a large
mass particle (wrapped D-brane), and can be inserted at different
points on the AdS sphere. Schwarzschild time runs in opposite
directions in the two asymptotic regions, so the observer in region
I can prepare an in-state of K particles,  send the beam through
the horizon, and then the other observer can observe an out-state
of $K'$ particles, assuming that we have analytically continued so
the bounce geodesic dominates. If the boundary times of these
operators are all near $t_c$, the geodesics are almost null and
hence almost on shell. A ``meta-observer" could in principle
compare the data of these two observers and compute a ``meta
S-matrix."  This quantity would seem to contain significant
information about the fate of organized matter as it approaches
the singularity.

But from the point of view of a global observer using Kruskal time,
these correlators correspond to particle-antiparticle annihilation
near the singularity, and the correlator is a kind of vacuum
persistence amplitude.  Its interpretation is less clear.

A question one typically asks about singularities in consistent
physical theories is what resolves them.  Here the black hole
singularity seems deeply connected to the boundary $t_c$
singularity in the analytically continued CFT amplitudes.  Perhaps
this $t_c$ singularity is smoothed out in the full CFT with finite
$\lambda $ and $g_s$.  Of course the difficulties described in the
previous section in extracting finite $g_s$ information make it
possible that this question can only be addressed in some double
scaling limit.

The definition of gauge theory correlators by analytic
continuation from the Euclidean thermal theory must determine some
boundary conditions for bulk fields at the black hole singularity.
Some work has been done on this question in the $d=3$ case
\IchinoseRG, but $d>3$ is technically more difficult.  We should
be able to address this issue by tying geodesic techniques to WKB
approximations of the wave equation.   This may be connected to
the possibility of defining different black hole systems by
starting with different initial states than the Hartle-Hawking
state.

One important tool for analyzing the boundary  CFT is to go to
small 't Hooft coupling $\lambda$ (at $N = \infty$) where weak
coupling perturbation theory is valid.   But, as mentioned in
Section 4.2, at weak coupling there is no sign of the branch cut
signalling the metastable phase on which we concentrate.  As noted
above it is not difficult to imagine $l_s$ corrections to the
metric and other supergravity fields that will remove the branch
point and bouncing geodesic.   It would be very useful to have an
understanding of this from the gauge theory side.


\vskip 1cm

\centerline{\bf Acknowledgements} We would like to thank Ben
Craps, Ben Freivogel, Gary Horowitz, Jared Kaplan, Per Kraus, David Kutasov,
Emil Martinec, Rob Myers, Mukund Rangamani, Simon Ross,
Andrei Starinets, and Lenny Susskind for helpful
discussions.  This work is supported in part by  NSF grant
PHY-9870115 and by the Stanford Institute for Theoretical Physics.
Matthew Kleban is the Mellam Family Foundation Graduate Fellow.

\appendix{A}{Kruskal coordinates}

We have motivated the non-square nature of the Penrose diagram by
studying radial null geodesics. Since the Schwarzschild
coordinates do not cover the whole spacetime,
 we have used complexified
coordinates in order to consider the globally extended spacetime.
However, for the purposes of understanding the global structure,
it is more desirable to pass to real, globally-defined
coordinates. The most convenient ones to use are the Kruskal
coordinates, which we now present.  This will allow us to see the
behavior of null geodesics, without having to use complexified
coordinates and to deal with coordinate singularities at the
horizons.

Many of the issues discussed in this and the next Appendix have previously been addressed
in a general context in \KloschQV\ .

Suppressing the angular directions, the metric \metric\ is
\eqn\metricrt{ ds^2 = - f(r) \, dt^2 + {dr^2 \over f(r)}} with $f
= r^2 - {1 \over r^2}$.  Now define the Tortoise coordinate
\eqn\rstar{ \rs = \int_0^{r} {dr'\over f(r')} + C} with C an
integration constant we will choose later, so as to make the
Kruskal coordinates everywhere real. Letting $u = t - \rs$, $v = t
+ \rs$, we have $ds^2 = -f \, du \, dv$. This is singular at the
horizon since the determinant vanishes; but
we can perform a further coordinate transformation $U = - e^{-2 u
}$, $V = e^{2 v}$ which will cast the metric into an extendible
form.  In these new coordinates, with $f(r)$ and $\rs(r)$ being
implicitly functions of $U$ and $V$, we have \eqn\kmetric{ ds^2 =
- { f \over 4} \, e^{-4 \rs} \, dU \, dV }
Letting $U = T-X$ and $V = T+X$, we can write this in a more
familiar form $ds^2 = g(T,X) \, (-dT^2 + dX^2)$. Note that by our
choice of
coordinates $U(u)$ and $V(v)$, the factor $g(T,X)$ in front of 
$(-dT^2 + dX^2)$ is nonsingular at the horizon: although $f$ by
itself vanishes at the horizon, the $e^{-4 \rs}$ factor cancels
this, so that the conformal factor $g$ remains finite across the
horizon. Therefore, we can extend $(U,V)$ from $U<0,V>0$, to $U,V
\in (-\infty, \infty)$, subject to the constraint $0 < r <
\infty$.

We easily derive the transformation laws\foot{
For general functions $f(r)$ in \metricrt\ with simple zeros
at $r= \rh$, we would have
$T^2 - X^2 = - e^{f'(\rh) \, \rs(r)}$ and
$2 \tanh^{-1} {T \over X} = f'(\rh) \, t$.
}
 between the old and the
new coordinates: \eqn\transa{ - e^{4 \, \rs(r)} = T^2 - X^2}
\eqn\transb{ 2 \, t = \tanh^{-1} {T \over X}} Note that $\rs$
acquires an imaginary part as the integral passes through the pole
at $1$. It is correspondingly convenient to define $C = {i \, \pi
\over 4}$, which makes $\rs$ real outside the horizon. More
specifically, we have
 $$ e^{4 \, \rs} = \( {r-1 \over r+ 1} \)
e^{2 \, \tan^{-1} r}$$

Using \transa, we can now read off the curves of constant $X^2 -
T^2$ labeling a constant $r$ surface.  In particular, it is easy
to see that the singularity ($r=0$) corresponds to $T^2 - X^2 =
1$, while the boundary ($ r \to \infty$) of \sads\ is given by
$X^2 - T^2 = e^{\pi}$.  Note that each of these equations is
satisfied by two disjoint curves, related to each other by $X \to
-X,\ T \to -T$. The full spacetime is then bounded by the four
hyperbolas in Kruskal coordinates, \eqn\STbound{ -e^{\pi} < T^2 -
X^2 = U \, V < 1} where the lower bound corresponds to the
asymptotic (timelike) boundary, and the upper bound to the
(spacelike) singularity. This is to be contrasted with the
corresponding situation for the $d=3$ BTZ spacetime, where a
similar analysis would yield $-1 < T^2 - X^2 = U \, V < 1$.

\ifig\figKrusk{Kruskal diagrams for \sads, with a) $d=3$ and b)
$d>3$. Radial null curves lie at 45 degrees; the horizons are
represented by the diagonal dashed lines.  The thick (spacelike)
hyperbolas correspond to the singularity, the timelike ones to the
boundary. The crucial difference between three and higher
dimensions is that in the former case, the hyperbolas are the same
distance from the origin, which translates into the geometric
property that any radial null curve bouncing around the diagram
ends up at the same point it started at. In higher dimensions this
doesn't happen.} {\epsfxsize=14cm \epsfysize=7.35cm
\epsfbox{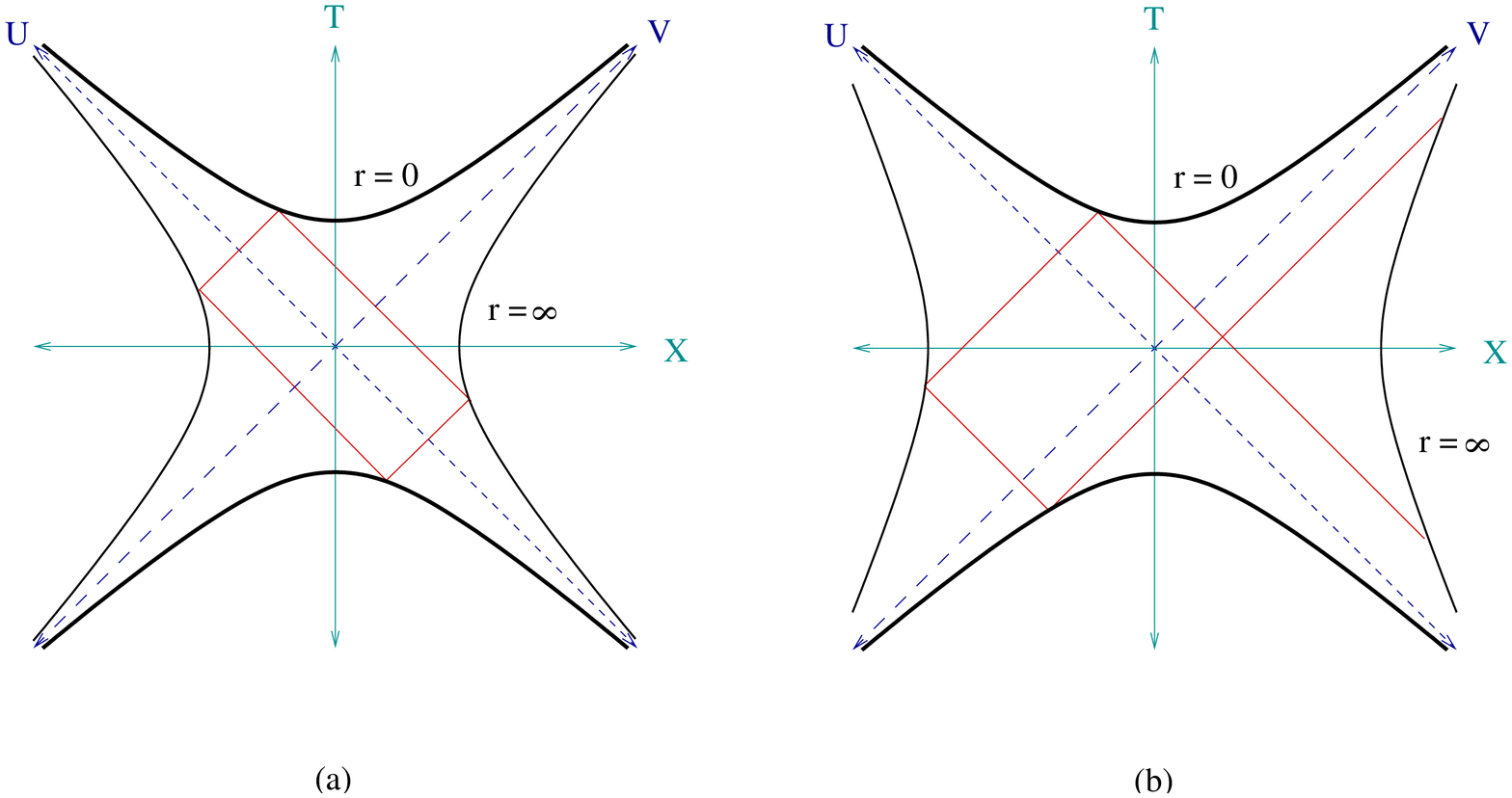}}

\figKrusk\ illustrates the contrast between the Kruskal diagram in
$d=3$ and $d>3$ dimensions.   For the 3 dimensional BTZ spacetime,
sketched in \figKrusk a, the singularities and boundaries
intersect the Kruskal coordinate axes at the same distance. On the
other hand, in all higher dimensions, the Kruskal diagram is as
shown in \figKrusk b, where the singularities come closer to the
center of the diagram than the boundaries. As we just computed,
for the $d=5$ large black hole, the ratio of these distances is
$e^{\pi} > 1$.

This immediately implies that there cannot be a radial null
geodesic starting on the boundary at $t=0$ (which implies $T=0$)
and hitting the singularity in the middle ($X=0$). A more
invariant way to say this is by considering a radial null geodesic
(or a sequence of geodesics), which ``bounce around\foot{
Specifically, starting from any point on the right boundary,
consider a future-directed ingoing null geodesic; from where this
hits the future singularity, draw the past-directed outgoing null
curve which eventually hits the left boundary; then  the
past-directed ingoing null curve which hits the past singularity;
and finally the future-directed outgoing null curve which comes
back to the right boundary again.} the diagram'', as shown by the
thin solid diagonal lines in \figKrusk. It is easy to see that if
the hyperbolas intersect at the same distance, as in \figKrusk a,
then such a sequence of null curves ends up at exactly the same
point where it started; otherwise, if the hypebolas intersect at
different distances as in \figKrusk b, the geodesics end at a
different point.  Furthermore, if the boundaries are further than
the singularities, these curves intersect inside the spacetime. As
we will discuss in Appendix B, this simple geometrical property
proves, in a more invariant way, that the Penrose diagram of
\sads\ cannot be drawn as a square for $d>3$, but can for $d=3$.

So far, we have presented our results only in the large black hole
limit. One may wonder whether for black holes comparable to (or
much smaller than) the AdS radius, this effect does not go away.
In fact, as we now explain, it gets larger. The smallest value
that the magnitude of the lower bound in
 \STbound\ can have occurs in
the large mass limit, where this goes to $X^2 - T^2 = e^\pi$. For
smaller black holes this value increases, and in fact gets
arbitrarily large  in the small mass limit: the boundary is then given
by $X^2 - T^2 = e^{\pi {R \over \rh}}$ with $\rh$ and $R$ denoting
the black hole and AdS radius, respectively. As we discuss next,
on a Penrose diagram, such as sketched in \figPdg b, the smaller
black holes would have the singularities more bowed in (using the
same conformal rescaling), or equivalently the boundaries more
bowed out in \figPdg c.  But this is exactly what we would expect
if we naively cut-off the asymptotically flat Schwarzschild
Penrose diagram at larger and larger distances $R$ (effectively
corresponding to smaller and smaller black holes).

\appendix{B}{Penrose diagram}

We have already discussed the qualitative features of the causal
structure of \sads\ black holes, but let us now confirm this by
finding the explicit Penrose diagrams for these spacetimes. This
will be achieved by finding appropriate transformation $V \to
\vv(V)$ and $U \to \uu(U)$ of the Kruskal coordinates $(V,U)$
(this guarantees that radial null geodesics will be 45 degree
lines), such that the boundaries of the spacetime now lie at
finite coordinate distance.

For example, letting $V = \tan {\vv \over 2}$ and $U = \tan {\uu
\over 2}$ compactifies the spacetime into a region inside $\vv,\uu
\in (-\pi,\pi)$. Letting $\vv = \tau + \rho$ and $\uu = \tau -
\rho$, we can now check that the singularities $U \, V = 1$ are
the straight lines $\tau = \pm {\pi \over 2}, \rho \in (- {\pi
\over 2}, {\pi \over 2})$. However, the boundaries, given by
$(\tan {\tau + \rho \over 2} \, \tan {\tau - \rho \over 2}) =
-e^{\pi}$, in the allowed region $\vv,\uu \in (-\pi,\pi)$, can be
explicitly checked to be ``bowed out''. On the other hand, if we
try to straighten out the boundary, e.g.\ by taking $V = e^{\pi
\over 2} \tan {\vv \over 2}$ and $U = e^{\pi \over 2} \tan {\uu
\over 2}$, so that $U \, V =  -e^{\pi}$ is given by $\rho = \pm
{\pi \over 2}, \tau \in (- {\pi \over 2}, {\pi \over 2})$, we find
that the singularity must be ``bowed in''. An exact plot
(generated by the latter transformation) was presented in
\figgeods b.

One may ask whether one can make a more clever choice of $\uu$ and
$\vv$ so as to straighten out {\it both} the singularity and the
boundary. After all, the usual lore that ``curves can be
straightened out by suitable conformal transformations'' should
apply here as well. But in fact, we see that the present problem
is sufficiently constrained so that this cannot happen.

\ifig\figquad{Most general form of Penrose diagram with timelike
boundaries and spacelike singularities being straight (thick lines
on the diagram), since the horizons (dashed 45 degree lines) must
join the opposite vertices.  A null ray bouncing around (solid
thin line) must come back to the same point.} {\epsfxsize=5cm
\epsfysize=6cm \epsfbox{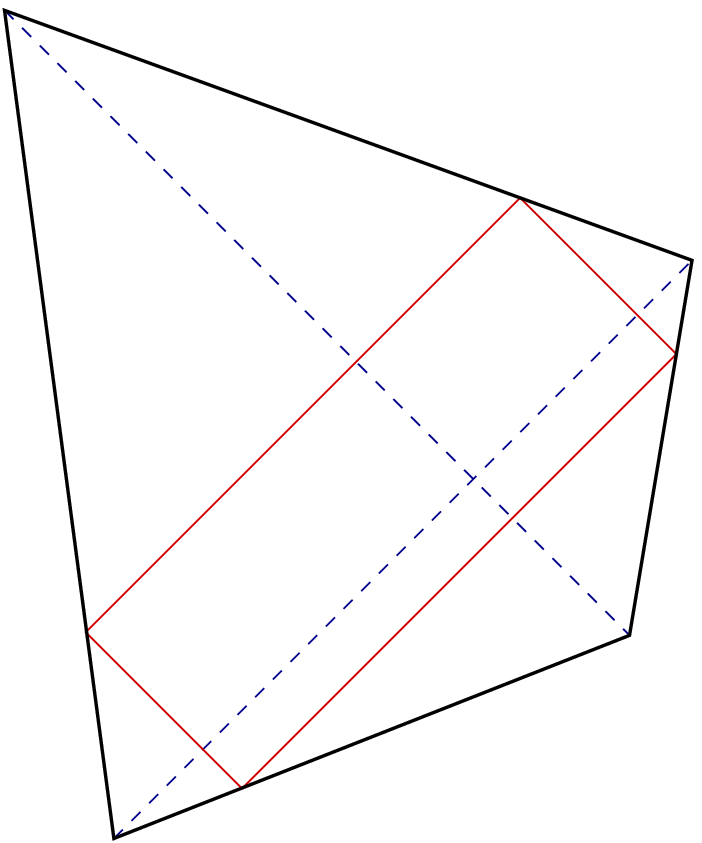}}

We can make a general geometric argument for why the singularities
and the boundaries cannot be drawn as straight lines, with the
whole spacetime filling some compact region on the diagram. The
first ingredient is to note that the horizons must be 45 degree
lines, and that the singularities and boundaries must ``meet'' at
these.\foot{By definition, horizons are the boundaries of the past
of infinity (so the boundary must touch the horizon on the Penrose
diagram), and any causal curve which enters the black hole must
end at the singularity (so similarly, the singularity must touch
the horizon in a Penrose diagram); and finally, the horizons
themselves cannot form part of the boundary of the spacetime, as
by definition they lie entirely in the interior.} (Note that this
is the reason the Penrose diagram could not be e.g.\ a rectangle,
which might naively seem consistent with the observed properties
of null geodesics.) The most general way the corresponding Penrose
diagram with straight singularities and boundaries could look is
sketched in \figquad. But one can easily show by a series of
similar-triangle arguments that a null curve bouncing around this
diagram (analogously to that sketched in \figKrusk) must come back
to the same point. This is in contradiction to the properties of
null geodesics, as we have demonstrated above, so the
singularities and boundaries cannot both be straight, if the
spacetime is to be mapped into a compact region.

The last requirement is very important: we can easily construct
maps which make both singularities and the boundaries straight,
but at the cost of the ``center'' (where the two horizons
intersect) being mapped to infinity.  One such example is to let
$U = \uu$ and $V = 1/\vv$. The diagram of the spacetime would then
not be compact, and would not qualify as a Penrose diagram. In
retrospect, it is clear that one can't straighten out {\it any}
two curves by conformal transformation, without letting them go to
infinity. For example, consider two distinct curves which have the
same endpoints.  Whatever conformal transformation we apply, the
two curves will still have the same endpoints. But clearly, in the
flat geometry  there is only a single straight line between any
two points, so if the conformal transformation straightens out
both curves without making them go off to infinity, it must be
singular (i.e.\ non-invertible).

We should make one cautionary remark: we have shown that, keeping
the boundaries straight, the singularities must be bowed in.  The
amount of bowing, however, does depend on the conformal
transformation used.  In other words, the crucial geometrical
property we used was that radial null geodesics bouncing around
the diagram end up at a different point from which they started.
However, by a suitable conformal transformation, we can bring this
point to appear arbitrarily close to the starting point,
effectively making the Penrose diagram look almost, but not quite,
like a square.

\appendix{C}{Spacelike geodesics with angular momentum}

In this paper, we have been using only radial geodesics in \sads.
 For completeness, let us now consider geodesics to ones carrying
some ``angular'' momentum.\foot{ We will use notation appropriate
to finite-sized (spherically symmetric) \sads\ black hole; in the
large black hole limit, the angular momentum becomes the linear
momentum along one of the translationally invariant directions.}
 In this case, the
behaviour of spacelike geodesics is altered significantly. In
addition to timelike and radial components, the geodesics have an
angular component, described by the constant of motion
 $L = \phd \, r^2$.
The radial equation \spgeod\ is now modified to \eqn\spgeodL{
\rd^2 = E^2 + f(r) \, \( 1 - {L^2 \over r^2} \)} which describes a
1-dimensional motion of a particle of energy $E^2$ in the
effective potential \eqn\Veff{
V_{\rm eff}(r) = 
                 -r^2 + L^2 + {1 \over r^2} - {L^2 \over r^4}}
We see that rather than being repelled by the spacelike
singularity, the spacelike geodesics which penetrate sufficiently
close to the singularity are now attracted, and therefore
terminate at the singularity. Namely, a spacelike geodesic with
energy $E$ and angular momentum $L$ starting at $r= \infty$ can
propagate to $r = \ri$, where $\ri$ is now a function of both $E$
and $L$, given by the solution of the
equation \eqn\riL{
 E^2 + f(\ri) \( 1- {L^2 \over \ri^2} \) = 0 }
Note that for $E=0$, for any $L$, $\ri = 1$ as before (assuming
$L<1)$; however for $E>0$ the behaviour is more complicated, and
depends on $L$. For $L>1$, we see that $\ri > 1$ since otherwise
the LHS of \riL\ would be strictly positive.  This means that such
geodesics can never cross the horizon, and in particular can never
connect the two boundaries of \sads.  Thus they are not relevant
for us, and we will henceforth consider only geodesics with $L \le
1$. For $L < 1$, the geodesics must propagate past the horizon
(otherwise  the LHS of \riL\ would again remain positive), and
there are now two possibilities: for high enough energy, the
geodesic terminates at the singularity, whereas for energies below
this critical energy, $E < \Ec$, it reaches only down to $\ri > L
> 0$, and reemerges in the other asymptotic region. The dividing
energy $\Ec$ is of course given by the maximum value of the
effective potential in \Veff, $\Ec^2 = f(\ri) \, \( {L^2 \over
\ri^2} -1 \).$ Note that this is consistent with our previous
picture, since $\Ec \to \infty$ as $\ri \to 0$, i.e.\ as $L \to
0^+$. In the small $L$ limit, this behaves as $\Ec \sim {1 \over L
\, R }$.

Confining attention to $L \le 1$ and $E \le \Ec$, let us consider
the time $t_0$ at which a spacelike geodesic of given $E$ and $L$
has to start in order to be symmetric. By the same arguments as
before, this is given by \eqn\toEL{ t_0(E,L) + i \, \beta / 4 = -
\int_{\ri(E,L)}^{\infty} {E \over f(r) \, \sqrt{ E^2 + f(r) \( 1 -
{L^2 \over r^2} \)} } \, dr  } As before, $t_0$ vanishes for
$E=0$, and becomes more negative as $E$ increases. Thus, to see
the possible range of $t_0$ for various allowed values of $E$ and
$L$, it suffices to consider the part of the integral around $\ri$
for energies close to the saturating value $\Ec$. Since the
effective potential peaks at $\Ec$, and $\ri(E \to \Ec)$
approaches the $r$-value at which $V_{\rm eff}$ attains its
maximum, we can approximate the effective potential near its
maximum by $V_{\rm eff} = - f(r) \( 1 - {L^2 \over r^2} \) \approx
\Ec^2 - \mu^2 \, (r-\ri)^2$, where the coefficient $\mu^2$ is some
complicated function of $L$ and the spacetime parameters.
Substituting this back into \toEL, we see that for energies $E
\approx \Ec$, the integrand becomes $\Ec/f(r) \over (r-\ri)$.
Since $f(r)$ is well-behaved at $r = \ri$ for finite $L$, we see
that the integral \toEL\ has logarithmic divergence at $\ri$.
Thus, unlike in the $L=0$ case where $t_0$ was bounded by $\tc$
(cf.\ \tdiff), for $L>0$ this bound disappears, and $t_0$ can have
arbitrary values.

Let us now consider the behavior of the proper length along these
symmetric spacelike geodesics carrying angular momentum. This is
given by the generalization of \lengthE, namely \eqn\lengthEL{
\len = 2 \int_{\ri}^{\rinf} {dr \over \sqrt{E^2 + f(r) \( 1 - {L^2
\over r^2} \) } }} where $\rinf$ is the upper cutoff and $\ri$ is
given by solution of \riL. Since we already analyzed the behavior
as $E$ increases, for simplicity we will now consider the $E=0$
case, to see how increasing $L$ affects the proper length $\len$.
For $E=0$, $\ri = 1$ independently of $L$, so we clearly see that
increasing $L$ must increase the proper length $\len$ (since the
denominator of the integrand in \lengthEL\ decreases). In fact, as
$L \to 1$, it is easy to see that the length diverges
logarithmically at the near-horizon region.

\appendix{D}{Derivation of the geodesic time $t(E)$}

In this appendix we derive the formula for the boundary time $t_0$
as a function of $E$ for the case of the symmetric spacelike
geodesics. We first look at the infinitely massive black hole.
 The integral to be done is \tstartsp ,
\eqn\parta{-t_0=\int_{\ri}^{\infty} {E \over f(r) \, \sqrt{ E^2 +
f(r) }} \, dr}  where $\ri$ is the turning point and $f(r) = r^2 -
{1 \over r^2}$. Letting $r^2 = u$, $\gamma = \sqrt {1 + \quarter
E^4}$ and $u=v \, \gamma- \half E^2$, we obtain
\eqn\partb{-t_0=\int_{\ri^2}^{\infty} {\half E \, u \, du \over
(u^2 - 1) \sqrt {{(u + \half E^2)}^2 - (1 + \quarter E^4)}} =
\int_{1}^{\infty} {\half E \, (v \, \gamma - \half E^2) \, \gamma
\, dv \over ((\gamma \, v - \half E^2)^2 - 1) \, \gamma \, \sqrt
{v^2 - 1}}.}
 Now
let $v = \cosh x$. The integral becomes
\eqn\partc{-t_0=\half E
\int_{0}^{\infty} {(\gamma \cosh x - \half E^2) \, dx \over \gamma^2
\, v^2 - \, \gamma E^2 v + \quarter E^4 - 1} = {E \over 4 \gamma}
\int_0^\infty {(2v - {E^2 \over \gamma}) \, dx \over {v^2 - {E^2
\over \gamma} v + \quarter E^4 - 1}}}
Note that the numerator is
the derivative of the denominator with respect to $v$.  This makes
the decomposition into partial fractions easy, and the integral
becomes
\eqn\partd{- t_0={E \over 4 \gamma} \int_0^\infty {dx \({1
\over v - \alpha} + {1 \over v - \beta}\)}}
where $\alpha = {\half
E^2 + 1 \over \gamma}$ and $\beta = {\half E^2 - 1 \over \gamma}$
are the two roots of the denominator of the previous equation. The
integrals can now be done by substituting $w = e^x$ and writing
$\cosh x$ in terms of $w$. The answer is
\eqn\parte{-t_0={E \over 2
\gamma} \[ {1 \over \mu_\alpha - \nu_\alpha} \ln \({1 - \nu_\alpha
\over 1 - \mu_\alpha}\) + {1 \over \mu_\beta - \nu_\beta} \ln \({1
- \nu_\beta \over 1 - \mu_\beta}\)\]}
where $\mu_x = x + \sqrt{x^2
- 1}$ and $\nu_x = x - \sqrt{x^2 - 1}$. Using the actual
expressions for $\alpha$ and $\beta$ in terms of $E$, we can
simplify the answer to the following form:
\eqn\partf{-t_0=\quarter \ln
\({\half E^2 - E + 1 \over \sqrt{1 + \quarter E^4}}\) - {i \over
4} \ln \({- \half E^2 + i E + 1 \over \sqrt{1 + \quarter E^4}}\)}
Here we have subtracted the imaginary piece $i \beta/4$
corresponding to crossing one horizon.  One can check that this
gives $t_c$ for large $E$ and that it goes like $E^3$ for small
$E$. This shows that there are 3 branches corresponding to the
positive $t$ direction starting at $0$ on the complex $E$ plane.

Recall that the result \longeqn\ quoted in Section 3.2 pertained to
the time difference between two boundaries, so that for symmetric
geodesics, $t =- 2 \, t_0$ (defined to be real for real energies $E$).

The case of finite mass black hole  can also be done exactly, though
it is  computationally more involved.  In particular, following
the procedure outlined above for the metric \metric, we obtain
\eqn\tfinite{\eqalign{ -t_0(E) = {R^2 \over 2 \Rrr} \ \{
 \rh \, & \ln \( {(1+E^2) \, R^2 - 2 E \, R \, \rh + 2 \rh^2 \over
\sqrt{(1+E^2)^2 \, R^4 + 4 R^2 \, \rh^2 + 4 \rh^4}} \) \cr - i \,
\sqrt{\rh^2 + R^2} \, & \ln \( {(1-E^2) \, R^2 + 2 \, i \, E \, R
\, \sqrt{\rh^2 + R^2} + 2 \rh^2 \over \sqrt{(1+E^2)^2 \, R^4 + 4
R^2 \, \rh^2 + 4 \rh^4}} \) \} }} Note that for $\bar{t} = {\rh
\over R^2} t$ and $\bar{E} = {R \over \rh} E$, in the ${R \over
\rh} \to 0$ limit, $-\bar{t}_0(\bar{E})$ reduces to \partf\
with barred $t$ and $E$.

Expanding around $E \approx \infty$ yields $t_c = -{1 \over 4 T} {
\sqrt{\rh^2 + R^2} \over \rh }$ which can be easily verified by
considering radial null geodesic as in \tra.

Expanding around $E \approx 0$ yields \eqn\Eseries{ -t_0(E)
\approx  {R^5 \over \Rrr^2} \, E + {R^5 \, (2 \rh^2 - R^2)^2 \over
3 \, \Rrr^4} \, E^3 + \CO({E^5}) .}
In passing we note
the intriguing fact that the cubic term flips sign when $\rh = {R
\over \sqrt{2}}$, which is exactly the same as the transition point when
the specific heat becomes negative---that is, at the Hawking-Page
transition \HawkingDH.  Although the physical significance is not
clear, it is rather suggestive that these small black holes (which
are no longer approximately thermal states in the CFT) have a vastly
richer geodesic structure.

The important point, used in Section 3.3, is that $t(E)$ now has
the form \eqn\finalresult{t(E) \sim  E^3 - a E} with $a<0$, so
that the cubic degeneracy is resolved for Euclidean time.

\appendix{E}{Analytic structure of $E(t)$}

We have given above a formula for $t(E)$ in the infinite mass
black hole limit.  This formula defines a Riemann surface over the
$t$-plane, and it is amusing to examine its analytic structure.
Note that because of formula $\lengthEr$ the analytic structure of
$E(t)$ is closely related to that of the large $m$ correlation
function (modulo extra branch cuts from the log in $\lengthEr$).
To start, we find the locations of all the branch points of
$E(t)$. These must correspond to values of $E$ for which $dt / dE$
vanishes, and a quick examination of $dt / dE$ shows that it
vanishes only at $E = 0, \infty$.

Now $t$ is periodic with period $i \beta= i \pi$. For $E(t)$, we
have first of all the coincident points pole at $\pi i / 2$ (with
$E \sim {1 \over t - \pi i / 2}$); our previous analysis shows
that we have a branch cut at $t=0$ (with $E \sim t^{\third}$), and
the $t_c$ pole at $\pi / 2$. In fact this pattern of branch cuts
and poles repeats itself, with $t^{\third}$ branch cuts
corresponding to points where $E=0$ and thus (looking at the
residues of the logs in $\longeqn$) occurring at $t = (p + iq) \pi
/ 2$ with $p + q$ even, and $1/t$ poles corresponding to points
where $E = \infty$ and occurring at $t = (p + iq) \pi / 2$ with $p
+ q$ odd. Of course, these poles and branch cuts only occur on
certain sheets of the Riemann surface. It is not too difficult to
understand the sheet structure; the only hard part is seeing what
happens at the branch points.  There we have $3$ sheets.  It turns
out that two of those (the non-$t_c$ branches) do not have any
singularities or branch cuts.  Moving far in Lorentzian time along
these branches corresponds to $E$ spiraling in to one of the
fourth roots of $-4$, and thus getting far away from $E = 0,
\infty$, where the poles and branch points occur.  Of course, the
non-$t_c$ branches are the physical ones, so in some sense all the
interesting analytic structure is very nonphysical.  One possible
interpretation for the string of $t_c$ poles is that they
correspond to geodesics bouncing off both the singularities and
the boundaries of \sads\ multiple times.  In order to bounce off a
singularity the geodesic has to be spacelike whereas to bounce off
the boundary of \sads\ it has to be timelike, so this idea is
speculative.  It is worth noting, however, that the poles on both
boundaries predicted above are precisely the locations where the
geodesic bounces.

\appendix{F}{Computing properties of the $t_c$ singularity}

In the text we argued that certain information about the black
hole singularity is encoded in boundary correlators.  In
particular there is information in a singularity at $t=t_c$ on a
secondary sheet of the two point correlator, which at $m= \infty$
has a third order branch point at $t=0$.  In this appendix we will
show that this singularity can be studied in a computationally
effective way, assuming that precise numerical information about
the boundary correlator as a function of Euclidean time on the
primary sheet is available.  Of course such information is not
available (for large $\lambda$) from direct gauge theory
computation at this time.  But we feel it is useful to show by
example how the process of analytic continuation can be made
concrete given well defined gauge theory results.

Specifically we assume that the a number of coefficients in the
Taylor expansion of the gauge theory  correlator $\langle \phi\phi
\rangle(t)$ are known. $L(t)= {-1 \over m} \ln \langle \phi\phi
\rangle(t)$
 has a smooth large $m$ limit so we consider its
expansion.  For convenience we expand around  euclidean antipodal
separation, $t=0$ even though $L(t)$ is branched there. The third
order character of this branch point is known from our theoretical
analysis or would be clear from numerical data. So we write our
expansion in terms of  $x = t^{2 \over 3}$. For Euclidean time
separations the large $m$ evaluation of $L(t)$ from a real
geodesic is certainly correct. So we can expand \lengthEr\ and
\longeqn\ to obtain model ``numerical" gauge theory data. Stopping
at order $x^{20}$ is sufficient to illustrate how things work.

\eqn\modser{\eqalign{L(x)= & ~~0.693147 - 1.36284\,x^2 -
   0.265333\,x^4  - 0.106447\,x^6 - 0.0454717\,x^8 \cr & -
   0.0199415\,x^{10}  - 0.0090257\,x^{12} -
   0.00421708\,x^{14} - 0.00202004\,x^{16}\cr & -
   0.000984367\,x^{18} - 0.000485493\,x^{20} + \ldots}}

The analytic structure discussed in Section 3.2 and Appendix E
implies that the $t_c$ singularity of $L$ on the secondary $t$
sheet will appear at  $x_c = t_c^{2/3} = 1.35128345.$ This
singularity has the leading behavior $L(x) \sim 2 \log(x-x_c)$.

To find this singularity numerically we use standard techniques.
First we differentiate \modser\ so that the leading singular
behavior is a simple pole, $L'(x) \sim 2/(x-x_c)$.  We then fit
\modser\ to a ratio of two polynomials $P$ and $Q$  of degree $M$
and $N$, $L'(x)\simeq
 P_M(x)/Q_N(x)$.  This is called a Pad{\'e}
approximant \Baker\ .   The convergence of such approximants is
nonuniform, so we select one more or less arbitrarily.

The $[M,N]=[7,12]$ Pad{\'e} has a pole at $x=1.35128349$ (compared
to $x_c= 1.35128345$) with a residue of $2.0028$ (compared to $2$)
. These results are probably anomalously good.  The $[9,10]$
approximant has a pole value of $1.35137$.

So in this limiting case a modest amount of ``numerical" data is
sufficient to study the singularity.   But it is important to
ascertain how stable this situation is. In particular, as
discussed in Section 4.3, finite $g_s$ limits our ability to take
$m$ arbitrarily large.  We know that at finite $m$ the branch cut
at $t=0$ is erased and we can no longer analytically continue to a
secondary sheet. As discussed in Section 3.5, a signal of this
erasure order one corrections to the order $k$ power series
coefficient when $k \sim m$.   Here we see that $k=20$  is
adequate to determine $L$, so we just need $m>>20$.  As also noted
in section 3.5 there will be an asymptotic series of $1/m$
corrections that are branched and can be analytically continued to
the $t_c$ singularity.  These contain interesting physics, as
explained in Section 4.  However to study the leading behavior
they must be handled in some way, perhaps by extrapolation in
$1/m$.  But this may not be done with complete precision so we
investigate our ability to still see the $t_c$ singularity with
some small $1/m$ correction remaining.   In the $x$ variable this
correction has a pole at $x=0$ and a higher order pole at $x=x_c$.
The correct procedure would be to expand around a point $t_0 \neq
0$ and Pad{\'e} approximate both $x=0$ and $x=x_c$ singularities.

Rather than go through this entire procedure we model the
phenomenon by adding the series expansion of a pole at $x={1 \over
2} $,~ $1 \over m(x-1/2)$ to the derivative of \modser.  This pole
is much closer to the expansion point $x=0$ than $x_c$ is so its
power series coefficients grow much faster than those in \modser.
The ratio of the coefficients at order $n$ is approximately ${1
\over m} (x_c/(1/2))^n ~ \sim {1 \over m} (2.7)^n $ At $n=19$ for
$m=100$ the $1/m$ correction is about $10^7$ times larger than the
main result!  Nonetheless the Pad{\'e} approximant can efficiently
separate out these different poles.   The $[7,12]$ approximant
(with $m=100$) yields a pole at $x=1.35169$ with residue $2.0086$,
as well as the pole at $x=1/2$.

As explained in Section 4.1 we expect that the high curvature
around the black hole singularity will create a leading $1/m$
correction of form ${1/(m(t-t_c)^2)}$.  The series expansion of
this correction will eventually dominate over the terms in
\modser.  The ratio of terms of order $n$ will be roughly ${1
\over m} n^2$. There exist techniques to disentangle such
confluent singularities in a Pad{\'e} approximant \Baker\ .  But
we will content ourselves with showing that this modest rate of
growth allows us to ignore $1/m$ corrections for $m \sim 10^4$.
The $[7,12]$ approximant with this addition gives a pole at
$x=1.35061$ with residue $1.990$. This analysis illustrates  how
the apparently rather abstract notions of the $ m \rightarrow
\infty$ limit and subsequent analytic continuation to a
singularity on a secondary sheet can be made computationally
explicit.  Of course the input data, Euclidean gauge
theory correlators, do not yet exist!

\listrefs

\end